\documentclass[twocolumn]{pasj00}
\usepackage{graphicx}

\begin{document}
\SetRunningHead{S.Akiba, M.Nakada, C.Yamaguchi, and K.Iwamoto}{Gravitational Wave Memory from the Relativistic Jet of GRBs}
%\Received{}%{yyyy/mm/dd}
%\Accepted{}%{yyyy/mm/dd}
%\Published{}%{yyyy/mm/dd}

\title{Gravitational Wave Memory from the Relativistic Jet 
\\ of Gamma-Ray Bursts}

%%% begin:list of authors
% Do NOT capitalize all letters in "textsc".
\author{Shota \textsc{Akiba}, Megumi \textsc{Nakada}, Chiyo \textsc{Yamaguchi}}
\and
\author{Koichi {\sc Iwamoto}}
\affil{Department of Physics, College of Science and Technology, Nihon University}
\email{iwamoto@phys.cst.nihon-u.ac.jp}
%%% end:list of authors

%%% Please use the following style in case that sorting by 
%%% affiliation is impossible. 
%
% \author{%
%   D-Firstname \textsc{D-Familyname}\altaffilmark{1}
%   E-Firstname \textsc{E-Familyname}\altaffilmark{1,2}
%   and
%   F-Firstname \textsc{F-Familyname}\altaffilmark{2}}
% \altaffiltext{1}{Address of Institute}
% \email{ddddd@xxx.xxx.xx.xx}
% \email{eeeee@xxx.xxx.xx.xx}
% \altaffiltext{2}{Address of Institute}

%% `\KeyWords{}' always has to be placed before `\maketitle'.
\KeyWords{Physical Data and Processes: gravitational waves -- stars: winds, outflows} %Do NOT move this preamble from here!

\maketitle

\begin{abstract}
The gravitational wave (GW) memory from a radiating and decelerating point mass is studied in detail.
It is found that for isotropic photon emission the memory generated from the photons is essentially 
the same with the memory from the point mass that radiated the photons so that
it is anti-beamed. On the other hand, for anisotropic 
emission the memory from the photons may have a non-vanishing amplitude even if it 
is seen with small viewing angles. In the decelerating phases of gamma-ray burst (GRB) jets the kinetic energy of the jet is converted into the energy of gamma-ray photons.
Then it would be possible to observe a variation in the GW memory associated with GRB jets
on the timescale of the gamma-ray emission if the emission is partially anisotropic.
Such an anisotropy in the gamma-ray emission has been suggested by the polarizations
detected in recent observations of GRBs. The GW memory from GRB jets
would provide clues to clarifying the geometry of the jets and the emission mechanism 
in GRBs. Thus it will be an interesting target for the next generation detectors of the GWs.  
\end{abstract}

\section{Introduction}

Gamma-ray bursts (GRBs) are considered to be relativistic jets with Lorentz
factors \( \gamma \sim 10^2 - 10^3 \). The energy radiated in gamma-rays for a single burst amounts to \(E_{\gamma} \sim 10^{50}-10^{52}\) erg (\cite{Frail2001}), which 
may be supplied by kinetic energy of the relativistic jets (e.g.,\cite{Piran1999}).
The origin of such energetic jets currently debated includes the 
coalescence of double neutron stars (NSs) in a binary as well as the collapsar, which is the collapse of a 
rotating massive star ending up with a system composed of a black hole (BH) and an accreting torus (\cite{Woosley1993}). 
It has long been considered that NS binaries and collapsing massive stars are the primary targets for the gravitational wave (GW) detection for missions such as 
LIGO, TAMA, Virgo, and LISA. In fact, some experiments have already begun to 
put constraints on the rates of such astrophysical phenomena(\cite{Akutsu2006}; \cite{Harry2010}). 
Soon after the discovery of the supernova-GRB association (\cite{Galama1998}),  a hydrodynamical simulation for the collapsar model was carried out by \citet{MacFadyen1999} and later it has been studied intensively by a number of numerical simulations (\cite{Nagataki2009}; \cite{Sekiguchi2011}; \cite{Ott2011} to list only a few of the recent works).
It is now established that the collapsar scenario is the most plausible model that we now know for long GRBs. 

Then GRBs  turn out to be major astrophysical sources of GWs. The mechanism of 
GW emission from GRBs is quite similar to that of the GWs from supernovae, which is mainly caused by asphericity in the matter motion 
in a bouncing core of collapsing massive stars and the anisotropic neutrino emission that emerges from the core (\cite{MJ1997}; \cite{Dimmelmeier2007}). In the collapsar scenario of GRBs, the neutrinos are emitted 
from an accretion torus and the total energy of neutrinos 
are estimated to be \(E_{\nu} \sim 10^{53}-10^{54}\) erg. Such GW signals have been studied analytically (\cite{Hiramatsu2005}; \cite{Suwa2009}) and by special relativistic magnetohydrodynamics simulations (\cite{Kotake2012}). 
The GW from anisotropic neutrino emissions comes up as a burst with memory, that is, 
a sharp rise followed by a steady value of the metric perturbation (\cite{Braginsky1987}).
In fact the GW memory has been considered for a variety of sources such as
relativistic fluids (\cite{Segalis2001}; \cite{Sago2004}) and neutrinos ( \cite{Hiramatsu2005}; \cite{Suwa2009}) in the context of the GW from GRBs. 
The maximum amplitude of the GW memory produced by 
a relativistic point source with energy \(E\) is given by

\begin{equation}
\Delta h = \frac{G}{c^4} \frac{4 E}{d} 
\sim 1.1 \times 10^{-22} \left( \frac{E}{10^{51} {\rm erg}}\right)
\left( \frac{d}{1 {\rm Mpc}}\right)^{-1}.
\end{equation}

\noindent
where \(d\) is the distance to the source.
Here we mean by a 'point source' a point mass or a single ray of photons/neutrinos, where the point mass is an approximation to treat a fluid element with a non-zero mass density.

\citet{Segalis2001} studied the GW memory from a  
relativistic jet of GRBs.
They derived the angular dependence of the GW amplitude for
the case of a point mass and found that the amplitude becomes quite small 
in the direction of jet propagation (anti-beaming effect). Another important finding 
is that the GW memory is polarized 
in the direction from the source to the line of sight on the transverse plane ( in the TT gauge ).
\citet{Sago2004} analyzed the GW memory from the accelerating phases of the GRB jets.  
They find that the finite size of the opening half-angle \(\Delta \theta\) 
of the jet leads to the reduction of the GW memory for viewing angles \(\theta_v\) smaller than \( \Delta \theta \), which means that the GW memory is hard to
detect simultaneously with GRBs. 
The total energy of 
neutrinos is two orders of magnitude larger than
in the case of relativistic jets of GRBs. However,  the
anisotropy in the neutrino emission is weak compared to GRB jets, 
resulting in the same consequence that it is not likely that we could observe 
the GW memory at the same time with GRB events (\cite{Hiramatsu2005}; \cite{Suwa2009}).

The amplitude of the GW memory expected for the relativistic jet of GRBs is much smaller than that expected for neutrinos from GRBs. However, a possible detection of the GWs characteristic for the jets would support the general view that GRBs are associated with relativistic jets and might resolve the issues
on the physical mechanism of the gamma-ray emission.
\citet{Sago2004} investigated the GW memory from relativistic jets based on the 'unified model' of GRBs , in which a bunch of multiple sub-jets launched from a central engine are assumed to be seen as GRBs, X-ray flushes (XRFs), and X-ray rich GRBs, depending on the viewing angle (\cite{Yamazaki2004}).    
They also pointed out that GRB jets may be GW sources in deci herz frequency bands, reflecting
the characteristic time scale of the central engine's activity, and thus will be
suitable targets for DECIGO and BBO (\cite{Seto2001}; \cite{Sago2004}).

It is thought that GRB phenomena detected by GW memory will be seen from off-axis 
with \( \theta_v > \gamma^{-1}\) because of the anti-beaming effect. Thus, such phenomena should actually be 
observed as XRFs rather than GRBs, or completely be missed, in electromagnetic observations. 
In the GRB phenomena, the relativistic jet will lose its kinetic energy and will be decelerated 
because of the gamma-ray emission. Then part of 
the memory carried initially by the jet will be transformed into the memory produced by electromagnetic 
radiation. Although 
the anti-beaming effect does not exist for the memory generated by a single ray of photons, photons are emitted 
toward a small but finite solid angle with a typical opening half-angle of \(\sim \gamma^{-1}\) so that the phase cancellation of the amplitude still occurs. However, 
if the emission is partially anisotropic, we
might be able to see a variation in the GW memory from GRB jets 
for a moderate range of viewing angles 
\(\theta_v \sim \gamma^{-1} \sim \Delta \theta \)
on the time scale of  gamma-ray emission. In fact, 
the polarizations detected in recent observations of GRBs suggest that the gamma-ray emission is partially anisotropic (\cite{Steele2009}; \cite{Yonetoku2011}). 
 
In this paper, we will present the results of a detailed study on the change of the 
GW memory in the decelerating relativistic shocks of GRBs. The paper is organized as follows. In \S 2 we briefly describe the 
formulation for calculating the gravitational waveform and focus on the
change of GW memory generated by a radiating and decelerating point mass.
In \S 3, we apply the result of \S 2 to realistic cases corresponding to GRB phenomena.
The finiteness of the opening half-angle of the jet is now taken into consideration. Then our analyses 
will be applied to the internal shock model of GRBs. We also present calculated waveforms for a specific model
that has appeared in the previous literature. 
Finally, the summary and discussion including the detectability are given in \S 4.

\section{Gravitational Wave Memory from a Radiating Point Mass}

We assume a flat background space-time with the Minkowski metric \( \eta_{\mu \nu} = (-1,+1,+1,+1)\) 
and restrict ourselves to dealing with the small 
perturbation \(h_{\mu \nu} = g_{\mu \nu}-\eta_{\mu \nu}\) of the metric \(g_{\mu \nu}\). 
We set \(c=G=1\) hereafter.
Then the linearlized Einstein equation reads

\begin{equation}
\left( -\frac{\partial^2}{\partial t^2}
+ \triangle \right) \bar{h}_{\mu \nu} = -16 \pi T_{\mu \nu}
\end{equation}

\noindent
under the Lorentz gauge condition \( \bar{h}^{\mu \nu}_{\hskip 0.2cm , \nu} = 0 \), where 
\(\bar{h}_{\mu \nu} \) is defined as 

\begin{equation}
\bar{h}_{\mu \nu} = h_{\mu \nu}-\frac{1}{2} \eta_{\mu \nu} h^{\lambda}_{\lambda},
\end{equation}

\noindent
and \(T_{\mu \nu}\) is the energy-momentum tensor.
For a point mass \(m\) with a world-line \(x^{\mu}(\tau)\) parametrized 
by the proper time \(\tau\), \(T^{\mu \nu}\) is given by
 
\begin{equation}
T^{\mu \nu} (x) = \int m u^{\mu}(\tau) u^{\nu}(\tau)
\delta^{(4)}(x-x(\tau)) d \tau.
\end{equation}

\noindent
where \(u^{\mu}(\tau) \equiv \frac{d x^{\mu}}{d \tau} = \gamma  ( 1, \beta \mathbf{n} )\) and \(\beta\), \(\mathbf{n}\) are the velocity, the unit vector proportional to the velocity of the point mass, respectively.

Following the notation in \citet{Segalis2001}, the retarded solution of equation (2) is written as

\begin{equation}
\bar{h}^{\mu \nu} (x) = 4m \frac{u^{\mu}(\tau_r) u^{\nu}(\tau_r)}{-u_{\lambda}(\tau_r) \cdot
(x-x(\tau_r))^{\lambda} },
\end{equation}

\noindent
where \(\tau_r\) is the retarded time
defined by the conditions

\begin{equation}
(x-x(\tau_r))^{\mu} \cdot (x-x(\tau_r))_{\mu} = 0, \hskip 1cm x^0-x^{0}(\tau_r) > 0
\end{equation}

The transverse-traceless (TT) part of \(h_{ij}\) defined by \(h_{ij}^{TT} = P_{ik} h_{kl} P_{lj}
-\frac{1}{2} P_{ij} h_{kl} P_{kl} \), where 
\(P_{ij}=\delta_{ij}-n'_i n'_j \) and \(n'_i\) is \(i-\)th component of the unit vector from the source to the observer,
has been derived to be

\begin{eqnarray}
h_+ \equiv h_{xx}^{TT} &=& - h_{yy}^{TT} = \frac{2 \gamma m \beta^2}{d} \frac{\sin^2 \theta}
{1-\beta \cos \theta} \cos 2 \phi, \nonumber \\
h_{\times} \equiv h_{xy}^{TT} &=& h_{yx}^{TT} = \frac{2 \gamma m \beta^2}{d} \frac{\sin^2 \theta}
{1-\beta \cos \theta} \sin 2 \phi, 
\end{eqnarray}

\noindent
in \citet{Segalis2001}. Here we choose \(z\) axis to be the line of sight. \(d\) is the distance to the source and \(\theta, \phi\) are
the polar and azimuthal angle in spherical coordinates, respectively, that indicate the direction of the jet.
It should be noted that \(\beta\) and \(\gamma\) 
are evaluated at the retarded time. It is found that the \(\theta\) dependent part of \(h_+\) and
\(h_{\times}\), which we denote \(h(\theta)\), has asymptotic forms

\begin{equation}
h(\theta) \equiv \frac{\sin^2 \theta}{1-\beta \cos \theta}
\simeq  \frac{2 \theta^2}{\gamma^{-2} + \theta^2}
\end{equation}

\noindent
for \(\gamma >> 1\) and \(\theta << 1 \) and

\begin{equation}
h(\theta) \simeq 1 + \cos \theta 
\end{equation}

\noindent
for \(\theta\) large compared to \(\gamma^{-1} \). This angular dependence shows that
the GW memory is diminished in the forward direction
within \(\theta < \gamma^{-1}\), which is called as "anti-beaming" (\cite{Segalis2001}; \cite{Sago2004}).

Now we turn to the case of photons.
Since there is no rest frame for a zero-mass particle,
the world-line is a null geodesic, being parametrized 
as \(x^{\mu}(x^0)\) by the coordinate time \(x^0\). Then the energy momentum tensor 
is expressed similarly as 

\begin{eqnarray}
T^{\mu \nu}(x) &=& E n^{\mu}(x^0) n^{\nu}(x^0) \delta^{(3)}(\mathbf{x}-\mathbf{x}(x^0)) \nonumber \\
&=& \int E n^{\mu}(t) n^{\nu}(t) \delta^{(4)}(x-x(t)) dt
\end{eqnarray}

\noindent
where \(x^0(t) = t \), \(E\) is the energy of the photon, \(n^{\mu} = ( 1, \mathbf{n} )\) is
the unit four-vector that is proportional to the photon's four momentum.

Comparing equations (4) and (10), we notice that the retarded solution for a zero-mass particle
is obtained straightforwardly from the one for a point mass with a simple replacement 
\(\gamma m \rightarrow E\), \(\beta \rightarrow 1\),
\(u^{\mu} \rightarrow n^{\mu}\). Then we have the TT part of the metric perturbation as follows.

\begin{eqnarray}
h_+ &=& \frac{2 E}{d} (1+\cos \theta ) \cos 2 \phi, \nonumber \\
h_{\times} &=& \frac{2 E}{d} (1+\cos \theta ) \sin 2 \phi.
\end{eqnarray}

This expression is consistent with the formula given in \citet{MJ1997}, which 
was derived for the GW memory from supernova neutrinos based on the work of \citet{Epstein1978}. 
It has already been suggested that
 the above replacement seems to be valid in \citet{Sago2004}.   

The GW memory from photons ( equation 11) has the same angular dependence
with the memory from a point mass for \(\theta > \gamma^{-1}\) (equation 9).  However, for
small \(\theta\), the anti-beaming is not seen for photons unlike the case of a point mass.   
Then, we might think of the idea that we would observe an increase in the GW memory, for
small viewing angles,
under the conversion of energy from a point mass into photons. However, 
photons are emitted from the point mass
into a finite solid angle with a typical opening half-angle of \(\sim \gamma^{-1}\),
which results in the same effect as if the jet has a finite opening half-angle. 
That is, the amplitude is likely to be canceled for small \(\theta\)  if the emission would be axisymmetric.

To calculate the GW memory, it is useful to define another frame 
\( (x',y',z')\), where \(y'\)-axis coincides with \(y\)-axis and
the \(x'(z')\)-axis is obtained by rotating the \(x(z)\)-axis around the \(y\) axis by an angle
\(\theta_v \) so that \(z'\) axis corresponds to the forward direction of 
the point mass or the symmetry axis of the distribution of the photon emission (figure 1). 
Thus \(\theta_v \) is equal to the viewing angle of the jet.
In changing coordinates of both frames into the spherical coordinates, 
the following relations hold among \(\theta, \phi\) and \(\theta', \phi'\). 

\begin{figure}
\hskip 0cm
\begin{center}
\includegraphics[width=8cm]{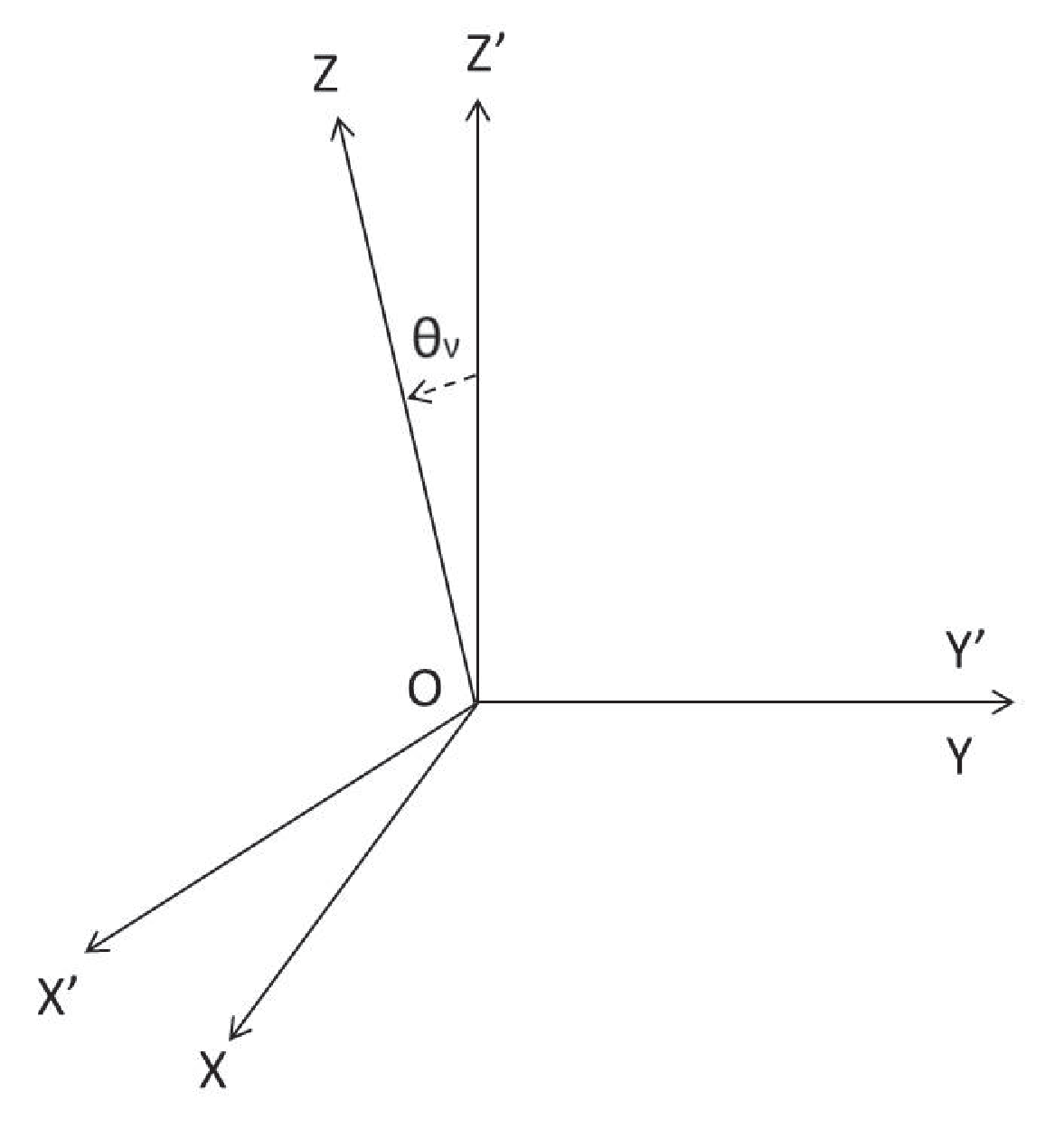}
    %%% \FigureFile(width,height){filename}
    \end{center}
  \caption{Two coordinate frames \((x,y,z)\) and \((x',y',z')\).
 The point mass (jet) moves in the \(z'\) direction and the \(z\) axis corresponds to the line of sight.  
The viewing angle is indicated by \(\theta_v\).}\label{fig:sample}
\end{figure}

\begin{eqnarray}
\sin \theta \cos \phi &=&
-\sin \theta_v \cos \theta' + \cos \theta_v \sin \theta' \cos \phi' \nonumber \\
\sin \theta \sin \phi &=&
\sin \theta' \sin \phi' \nonumber \\
\cos \theta &=& \cos \theta_v \cos \theta' + \sin \theta_v \sin \theta' \cos \phi' 
\end{eqnarray}
  
Then, the GW memory should be averaged over the angular distribution of photons as follows. 

\begin{eqnarray}
h(\theta_v) & \equiv & h_{+}(\theta_v) + i h_{\times}(\theta_v) 
= \frac{E}{d} < 2(1+\cos \theta) e^{2 i \phi} > \nonumber \\
&=& \frac{2E}{d} \int \int \sin \theta' d \theta' d \phi'
f(\theta',\phi') (1+ \cos \theta) e^{2 i \phi} 
\end{eqnarray}

\noindent
where \(\theta \) and \(\phi \) are related to \(\theta'\) and \(\phi'\) by equation (12), 
\(E\) is the total energy of the photons, and \(f(\theta',\phi')\) is the distribution function. The fraction of the energy of photons emitted into a solid angle \(d \Omega' = \sin \theta' d \theta' d \phi'\) is given by
\(f(\theta',\phi') d \Omega'\).  
Figure 2 shows a schematic view of the cone of emitted photons. 

We evaluate the GW memory
from photons for several cases of different angular distributions of the photon emission. 
First we assume an isotropic emission in the rest frame of the point mass for simplicity. 
Some GRBs have shown relatively high degrees of polarization in their early optical emission 
(e.g.,\cite{Steele2009}) and in gamma-ray emission itself (e.g.,\cite{Yonetoku2011}).  Such high degrees of 
polarizations may suggest the existence of ordered magnetic fields in the relativistic shock.
If the gamma-rays are caused by 
synchrotron radiation in ordered transverse magnetic fields (\cite{Granot2003}), 
the emission would tend to be focused into solid angles with a limited range of azimuthal angles so that the 
angular distribution will deviate from the axisymmetry.
To model such cases in a simplest way,  we study two cases. One is a maximally anisotropic
case where the photon emission is focused in the direction of a particular azimuthal angle. The other is the
case of synchrotron emission in the presence of ordered magnetic fields which has been considered in the
modeling of the polarization of the gamma-ray emission(\cite{Granot2003}; \cite{Toma2009}).   

\begin{figure}
  \begin{center}
\hskip 0cm
\includegraphics[width=8cm]{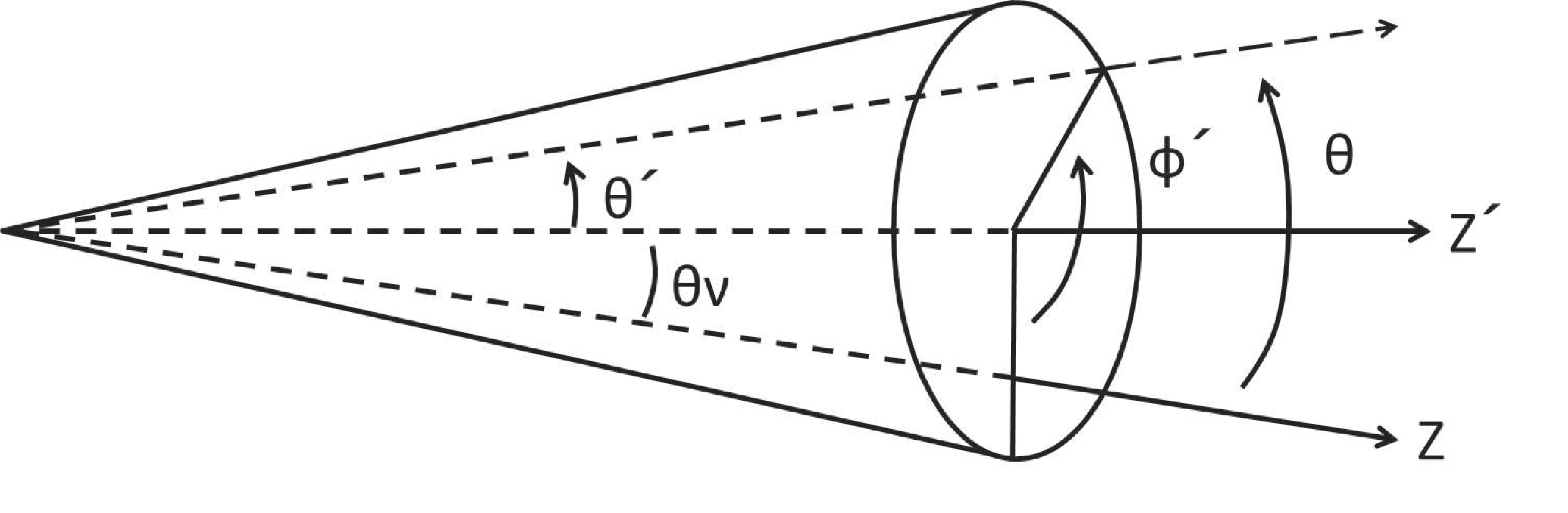}
    %%% \FigureFile(width,height){filename}
  \end{center}
  \caption{Schematic view of the cone of emitted photons}
\end{figure}

\subsection{ Case of Isotropic Emission}
 
If the photon emission is isotropic in the rest frame of the point mass, the angular distribution
function is given by
 
\begin{equation}
f(\theta',\phi') d \Omega' = \frac{1}{4 \pi} 
\frac{1}{\gamma^4 (1-\beta \cos \theta')^3} d \Omega'
\end{equation}

\noindent
and \(\beta, \gamma\) is the velocity and the Lorentz factor of the point mass, respectively (\cite{RPA}).
For \(\gamma >> 1\) the emission is beamed within small angles of  order \(\theta' \sim \gamma^{-1}\) so that
the amplitude depends only on \(\gamma \theta_v\). 
Figure 3 shows two polarization components \(h_+\) and \(h_{\times}\) of the GW memory 
from photons for isotropic emission generated by a point mass.
The normalized amplitude \(h/(E/d)\) is shown as a function of \(\gamma \theta_v\) (solid line). 
The \(h_{\times}\) component vanishes owing to the axisymmetry. 
We also plotted the memory
expected from a point mass with the same energy 
\(\gamma m = E \) (dotted line). Since \(h_+\) takes almost the same value 
with the memory from the point mass, it is hard to distinguish them in Figure 3.

\begin{figure}
\begin{center}
  \includegraphics[width=8cm]{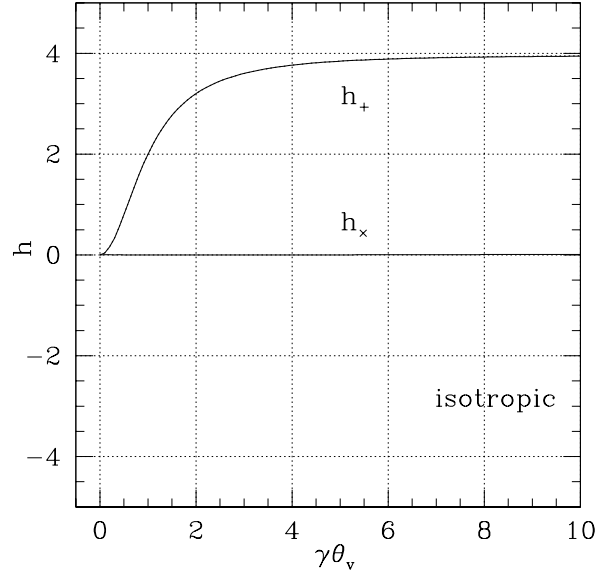}
\end{center}
\caption{
GW memory 
from photons with isotropic angular distribution of equation (14) radiated by a point mass 
with \(\gamma = 100\). 
The amplitudes normalized by \(E/d\) are shown as a function of the viewing angle \(\theta_v\) (solid lines). The \(h_{\times}\) component vanishes owing to the symmetry. 
The memory from a point mass with the same energy 
\(\gamma m = E \) is also plotted (dotted line), which takes almost the same values with \(h_+\) so that it is hard to distinguish them.}
\end{figure}

\subsection{Case of Maximally Anisotropic Emission}
  
Next we consider the second case in which the angular distribution is given by 

\begin{equation}
f(\theta',\phi') d \Omega' = \frac{1}{2} \frac{1}{\gamma^4 (1-\beta \cos \theta')^3} 
\delta(\phi'-\phi_0) d \Omega',
\end{equation}

\begin{figure*}
\hskip 0cm
\begin{minipage}[t]{0.95\textwidth}
\includegraphics[width=8cm]{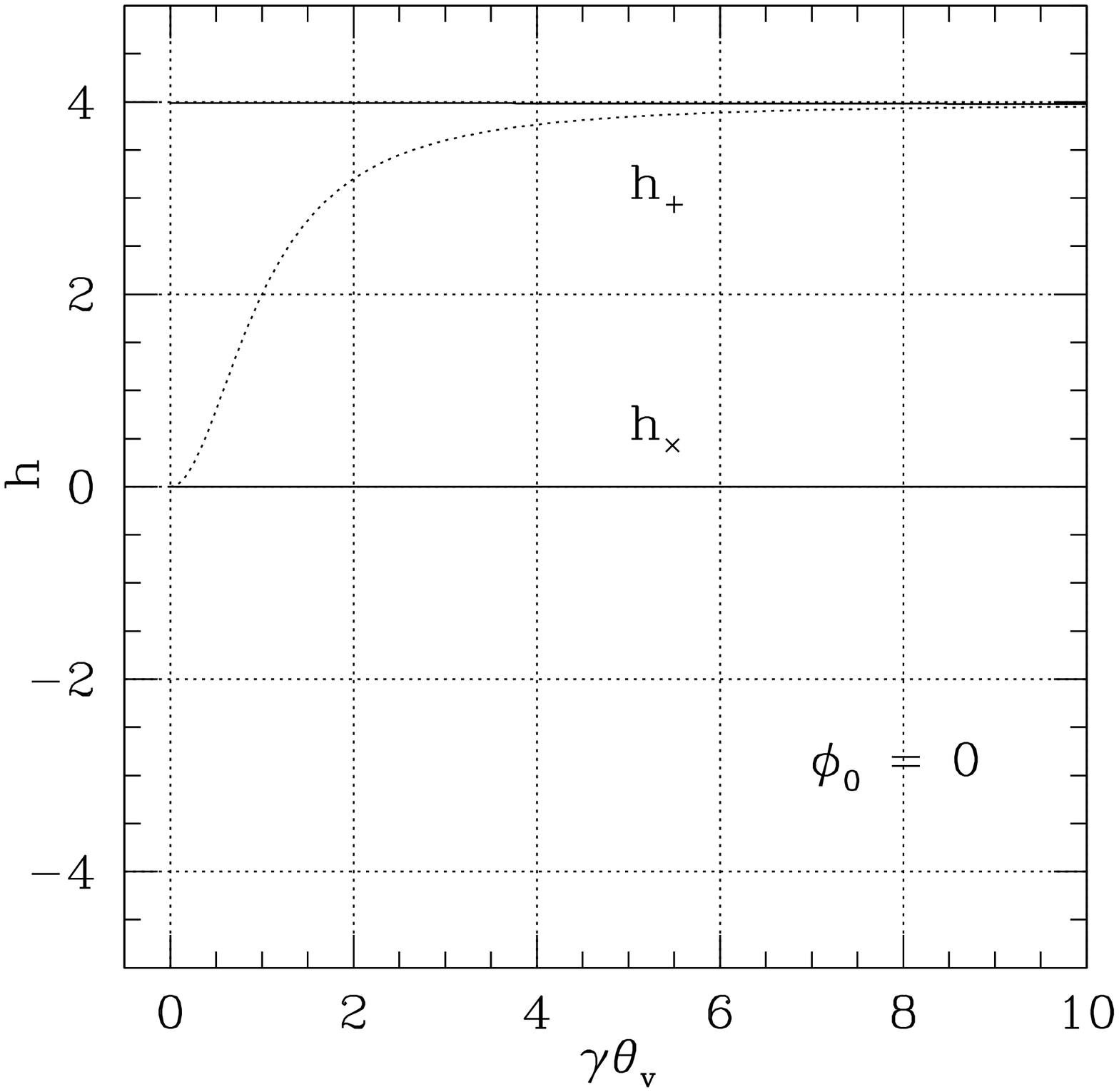}
\hskip 0cm
  \includegraphics[width=8cm]{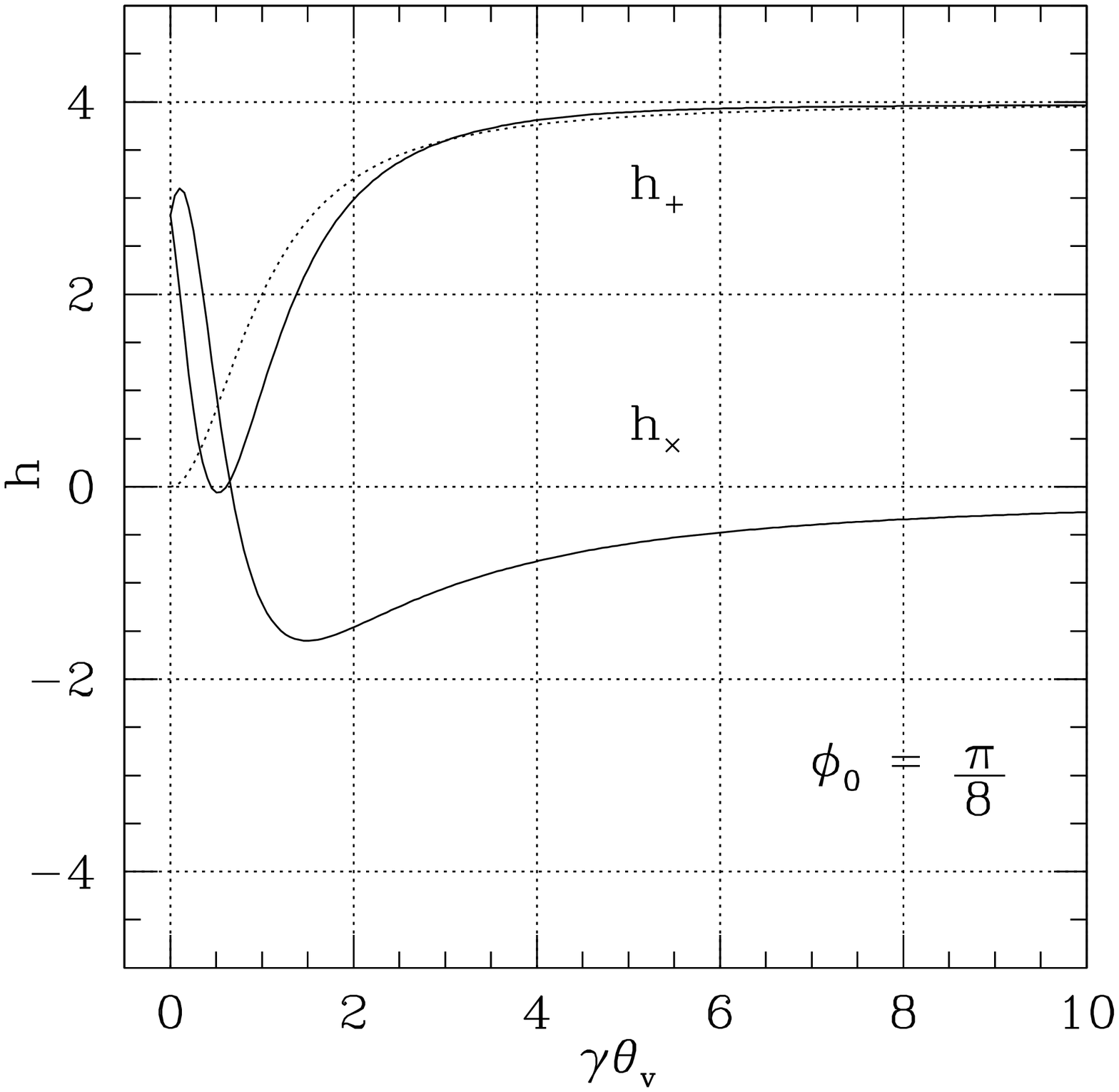}
 \end{minipage}
\hskip 0cm
\begin{minipage}[t]{0.95\textwidth}
   \includegraphics[width=8cm]{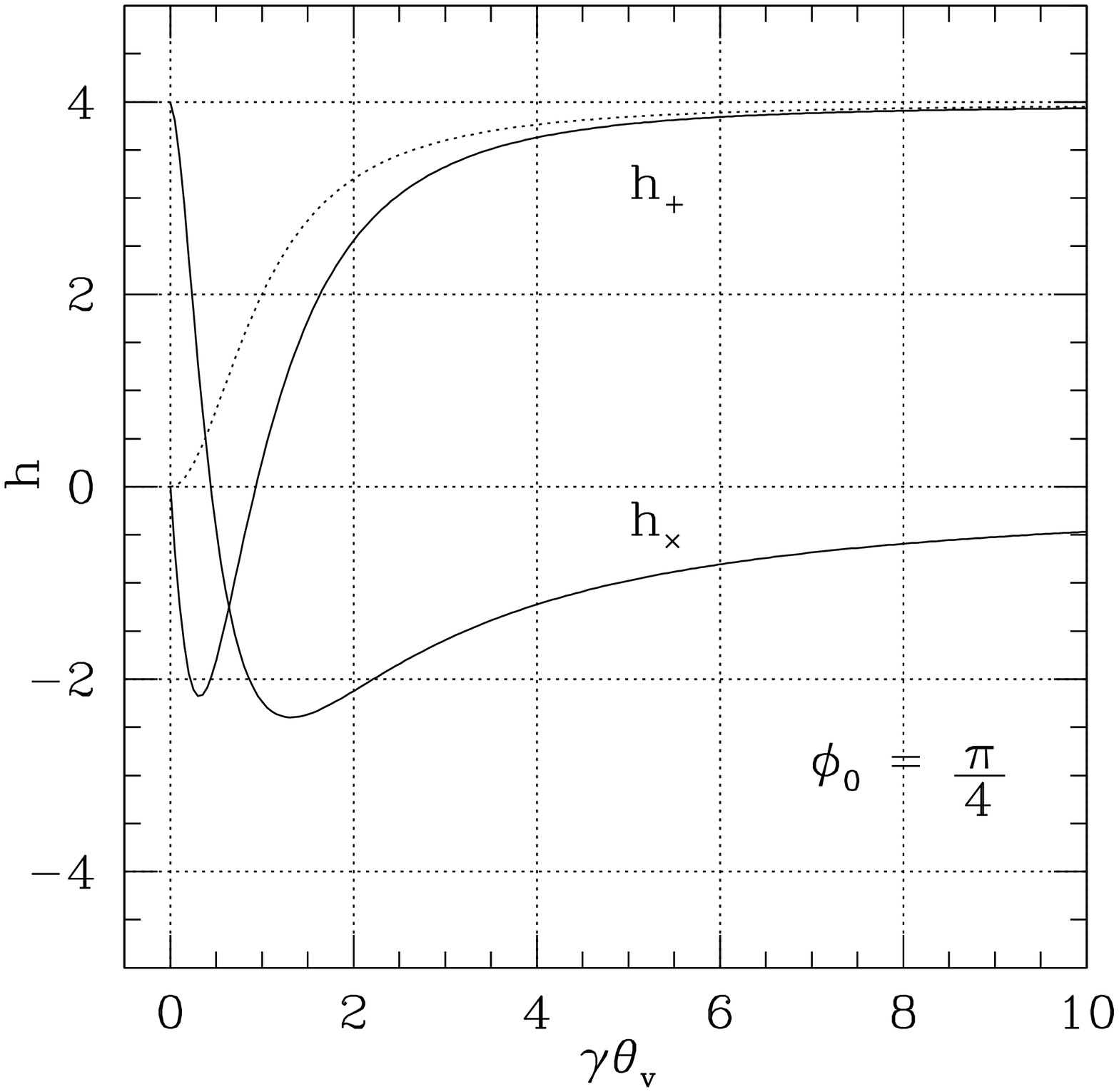}
\hskip 0cm
  \includegraphics[width=8cm]{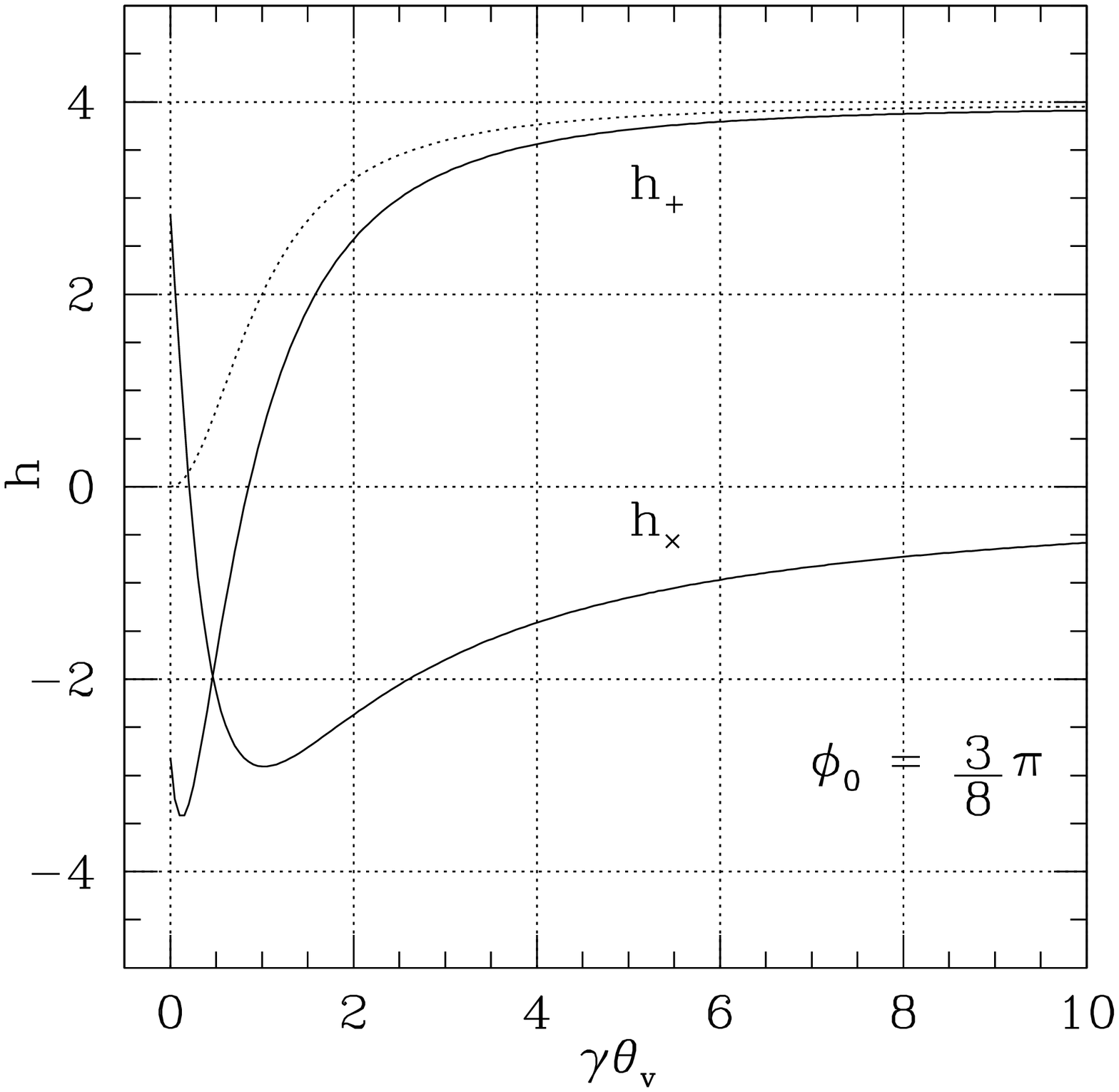}
 \end{minipage}
 \caption{ 
GW memory from photons with anisotropic angular distribution of equation (15)
normalized by \(E/d\) (case of maximally anisotropic emission).
Two polarization components \(h_+\) and \(h_{\times}\) (solid lines) are shown for 
\(\phi_0 = 0\) (top left), \(\frac{\pi}{8}\) (top right), 
\(\frac{\pi}{4}\) (bottom left), and \(\frac{3}{8} \pi\) (bottom right)
as a function of \(\gamma \theta_v \).
The amplitude from a point mass with the same energy \(\gamma m = E\) is shown for comparison (dotted line).}
\end{figure*}

\noindent
where \(\beta\) and \(\gamma\) is the velocity and the Lorentz factor of the point mass that radiates the photons.

For \(\theta_v = 0 \) the  
GW memory from photons with the angular distribution of equation (15)
can be evaluated analytically as

\begin{eqnarray}
h(0)  &=& \frac{2E (1+\beta)}{d} e^{2 i \phi_0} 
\end{eqnarray}

\noindent
For large \(\gamma \sim 10^2 - 10^3\) the amplitude of the 
GW memory takes a maximum value and the polarization is simply determined 
by the angle \(\phi_0\). 
   
For \(\theta_v \neq 0 \) we calculated the amplitudes averaged with equation (15) numerically. 
Figure 4 shows the two components of polarization 
\(h_+\) and \(h_{\times}\) calculated for
\(\phi_0 = 0, \frac{\pi}{8}, \frac{\pi}{4}, \frac{3}{8} \pi\)
as a function of \(\gamma \theta_v \). Figure 5 shows the same amplitudes but 
for \(\phi_0 = \frac{\pi}{2}, \frac{5}{8} \pi, \frac{3}{4} \pi, \frac{7}{8} \pi\).  
The amplitudes are normalized by \(E/d\) again.
The GW memory for a point mass with \(\gamma m = E\) is also shown 
with a dotted line for comparison. 
As \(\theta_v \) increases, \(h_+\) becomes a dominant component, approaching 
to the memory for a point mass asymptotically.

\begin{figure*}
\begin{minipage}[t]{.95\textwidth}
  \includegraphics[width=8cm]{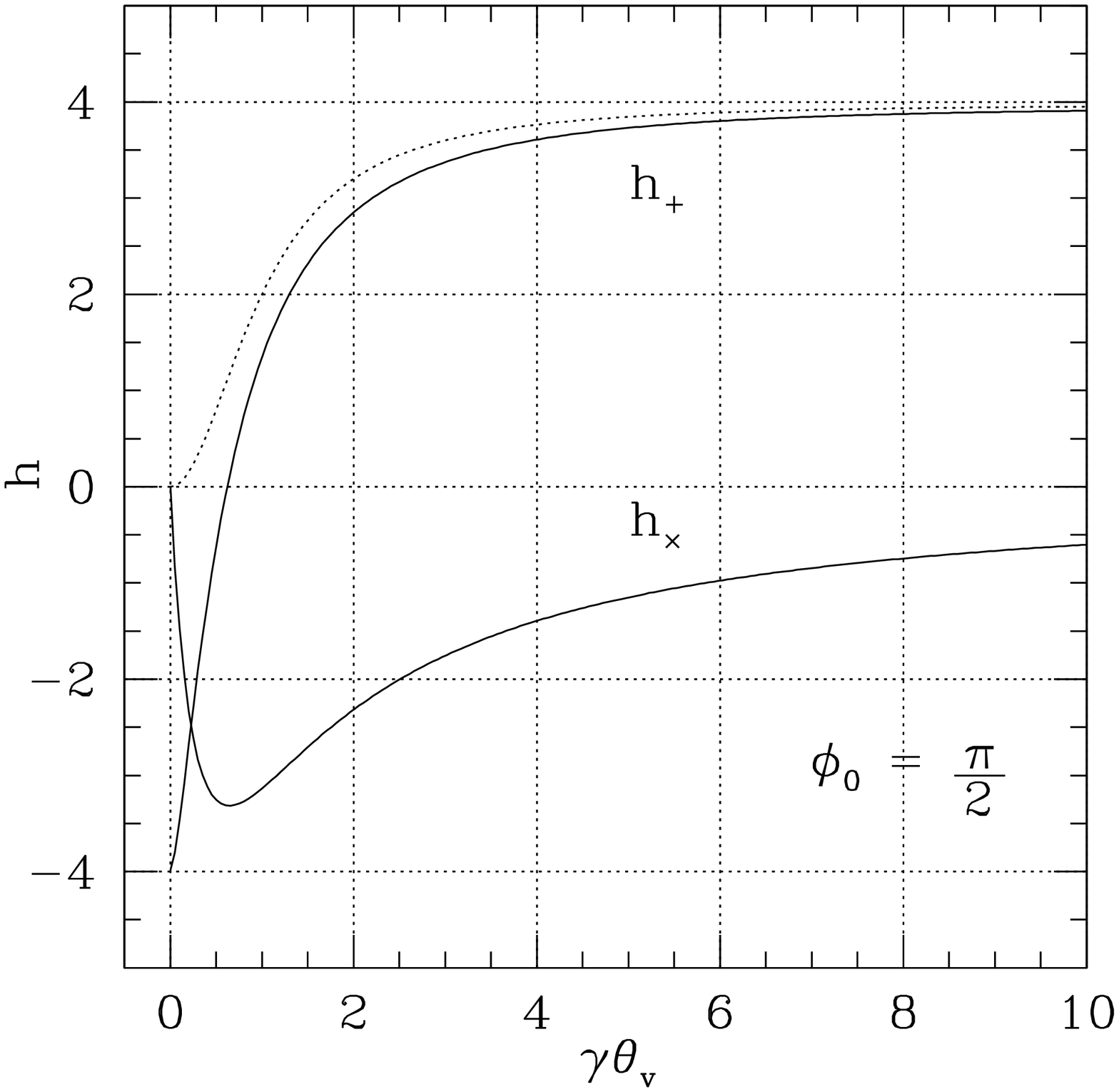}
 \hskip 0cm
  \includegraphics[width=8cm]{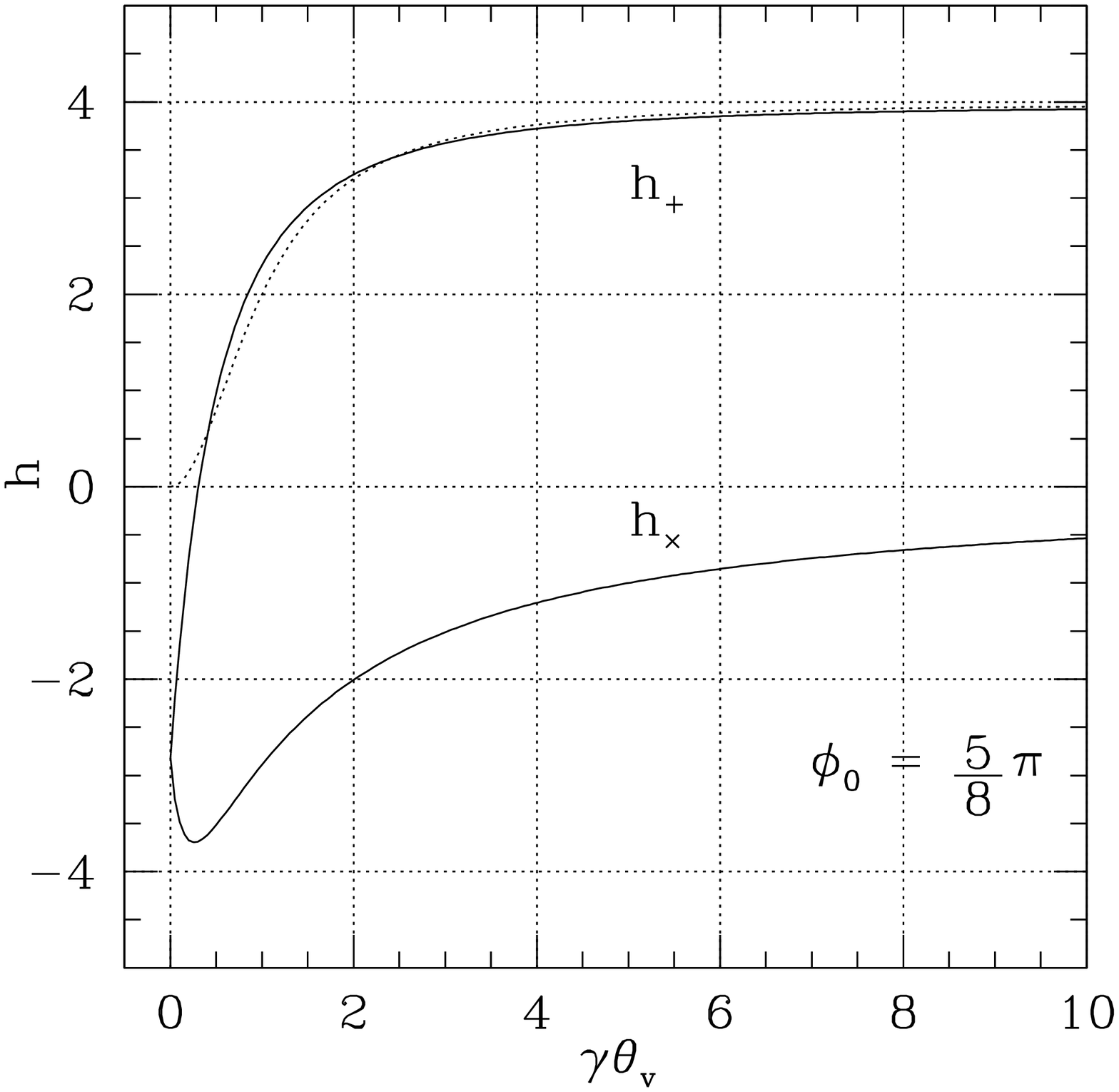}
 \end{minipage}
\hskip 0cm
\begin{minipage}[t]{.95\textwidth}
   \includegraphics[width=8cm]{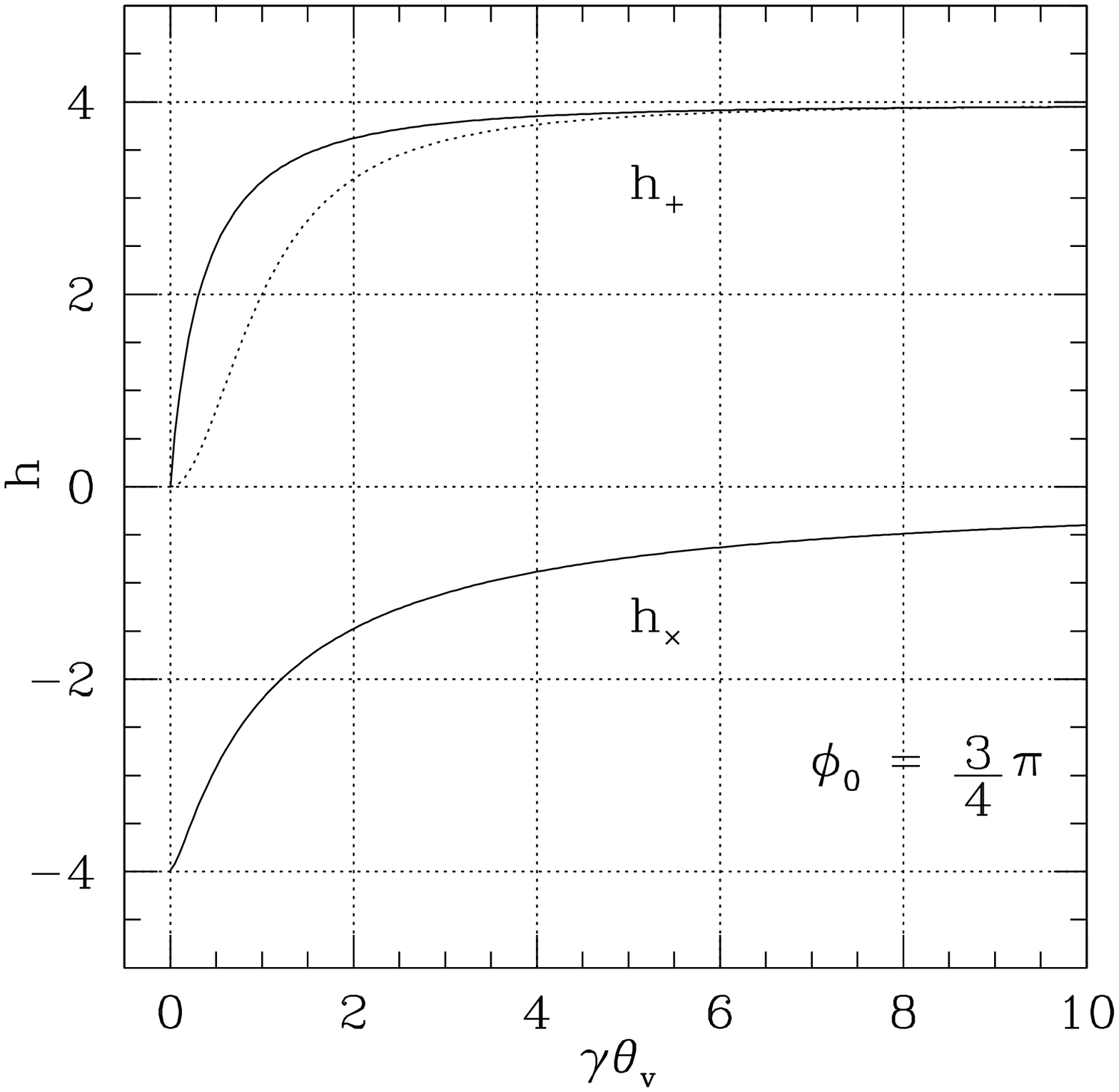}
 \hskip 0cm
  \includegraphics[width=8cm]{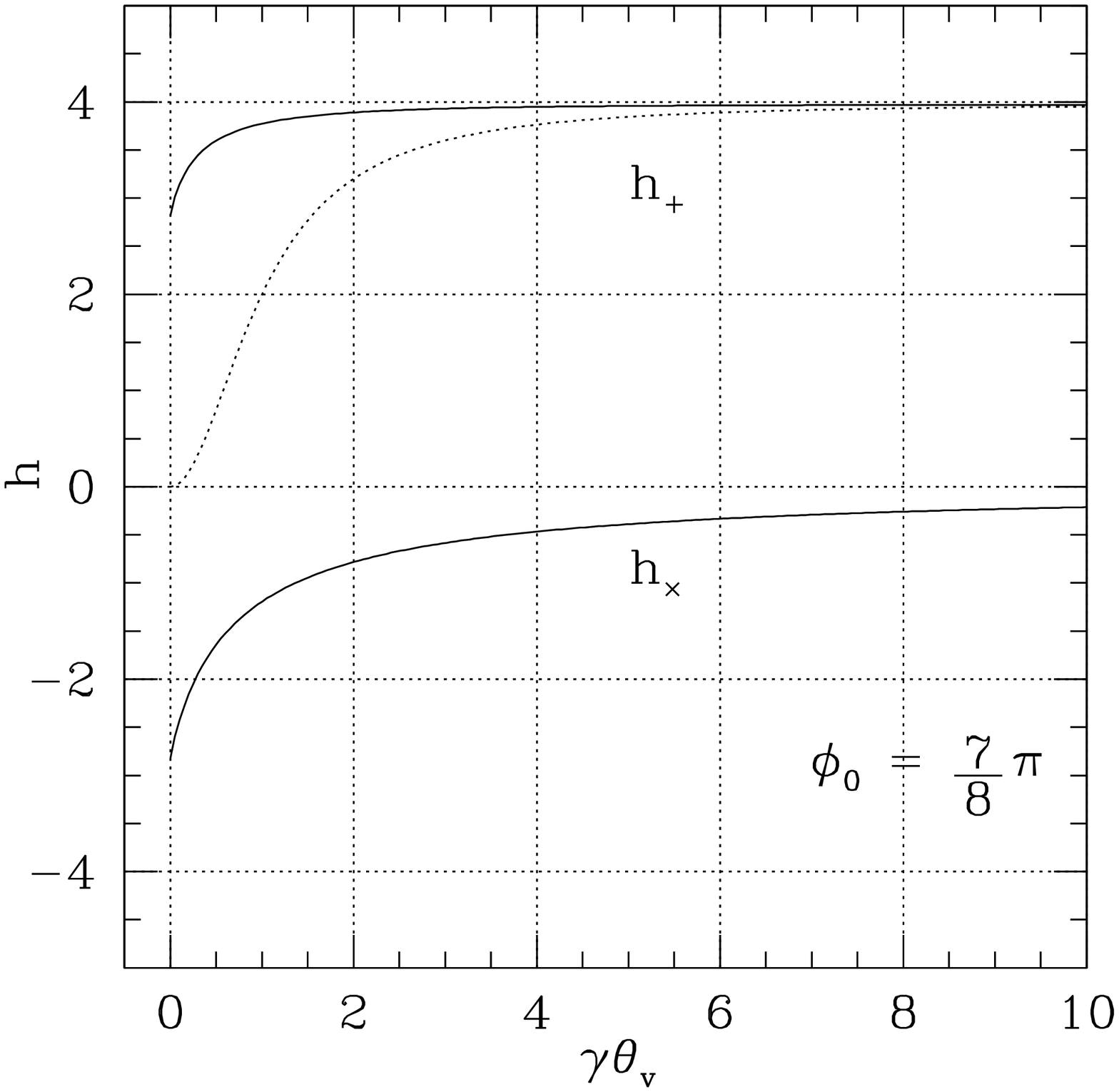}
 \end{minipage}
  \caption{The same with Fig.4 but 
for \(\phi_0 = \frac{\pi}{2} \) (top left), \(\frac{5}{8} \pi\) (top right), 
\(\frac{3}{4} \pi \) (bottom left), and \(\frac{7}{8} \pi\) (bottom right).
GW memory from photons with anisotropic angular distribution of equation (15)
normalized by \(E/d\) (case of maximally anisotropic emission).
The amplitude from a point mass with the same energy \(\gamma m = E\) is shown for comparison (dotted line).}
\end{figure*}

\subsection{Case of Synchrotron Emission with Ordered Magnetic Fields}

Here we consider synchrotron emission from electrons as a more physically-motivated model. 
We assume that an ordered magnetic field \( \mathbf{B} = (B \cos \phi_B, B \sin \phi_{B}, 0)\) 
is present ( Figure 6 ), the emitted photons have a simple power-law energy spectrum 
\(P_{\nu} \propto \nu^{-a}\), and that the electrons have an isotropic  
pitch angle distribution (\cite{Granot2003}).  
The photons emitted in the direction
\(\mathbf{n} = ( \sin \theta' \cos \phi', \sin \theta' \sin \phi', \cos \theta') \) have 
a pitch angle \(\xi\) such that 
\(\displaystyle{
\cos \xi = \frac{\mathbf{B} \cdot \mathbf{n}_c}{B}
}\), 
where \(\mathbf{n}_c\) represents the direction of the photons in the comoving frame,

\begin{eqnarray}
\mathbf{n}_c = ( \sin \theta_c \cos \phi', \sin \theta_c \sin \phi', \cos \theta_c ),
\end{eqnarray}

\noindent
and \(\theta_c\) is given by
\(\displaystyle{
\cos \theta_c = \frac{\cos \theta'-\beta}{1-\beta \cos \theta'}
}\), which reduces to
\(\displaystyle{ \cos \theta_c \simeq \frac{1-(\gamma \theta')^2}{1+(\gamma \theta')^2}}\)
for \(\gamma >> 1\)  and \(\theta' << 1\). 
We note that the magnetic field in the comoving frame is given by \(\gamma \mathbf{B}\) 
as a result of Lorentz transformation. 
A simple calculation yields 

\begin{eqnarray}
\cos \xi  &=& \sin \theta_c \cos \phi' \cos \phi_B + \sin \theta_c \sin \phi' \sin \phi_B \nonumber \\
&=& \sin \theta_c \cos (\phi'-\phi_B) 
\end{eqnarray}

Since the synchrotron power is proportional to \(( \sin \xi)^{a+1}\) (\cite{RPA}), 
the angular distribution is given by

\begin{eqnarray}
f(\theta',\phi')  d \Omega'= K (\sin \xi)^{a+1} \frac{1}{\gamma^4 (1-\beta \cos \theta')^3} d \Omega'
\end{eqnarray}

\noindent
where \(K\) is a normalization constant. We choose \(a=1\)  from a range of values considered  
in the literature (\cite{Granot2003}; \cite{Lazzati2006}; \cite{Toma2009}). For \(a=1\) the distribution function has a simple form and \(\displaystyle{
K = \frac{3}{8 \pi}}\). The GW memory depends on \(\phi_B\) as well as \(\theta_v\) so that we denote
the amplitude as \(h(\theta_v, \phi_B)\) in this case. 
For \(\theta_v=0\) the integration in equation (13) can be done analytically, where we find

\begin{eqnarray}
h(0, \phi_B) = -\frac{1}{4} \cdot \frac{2E(1+\beta)}{d} e^{2 i \phi_B}.
\end{eqnarray}
 
We see that \(\phi_0\) dependence of \(h(0)\) in equation (16) is reproduced by 
replacing \(\phi_B\) with \(\displaystyle{\phi_0-\frac{\pi}{2}}\)  in equation (20).
 For \(\theta_v \neq 0 \) we calculated the amplitudes averaged with equation (19) and \(a = 1\) 
numerically. 

\begin{figure}
  \begin{center}
\vskip 0.5cm
\includegraphics[width=8cm]{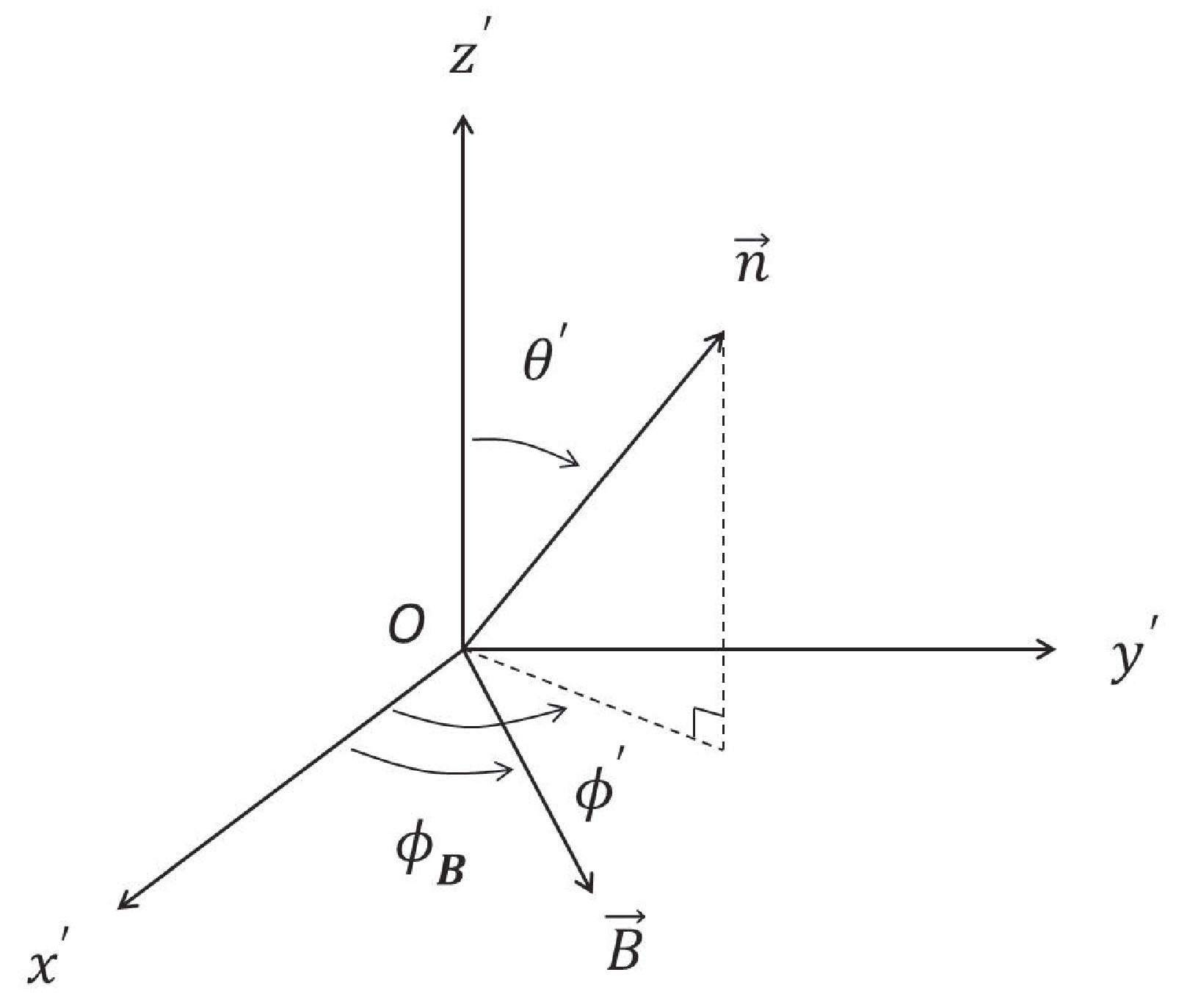}
    %%% \FigureFile(width,height){filename}
  \end{center}
  \caption{Geometry considered for the synchrotron emission from ordered magnetic fields.}
\end{figure}

Figure 7 shows the normalized amplitudes \(h(\theta_v, \phi_B) = h_+(\theta_v, \phi_B)+i h_{\times}(\theta_v, \phi_B) \) calculated for \(\phi_B = 0, \frac{\pi}{8}, \frac{\pi}{4}, \frac{3}{8} \pi\) as a function of \(\gamma \theta_v \). Figure 8 shows the same
for \(\phi_B = \frac{\pi}{2}, \frac{5}{8} \pi, \frac{3}{4} \pi, \frac{7}{8} \pi\).  
The GW memory for a point mass with \(\gamma m = E\) is also shown 
with a dotted line. The overall behavior is quite similar to the case of maximally anisotropic emission except that the amplitude for \(\theta_v = 0\) is reduced to one fourth of the maximum amplitude.    

\begin{figure*}
\hskip 0cm
\begin{minipage}[t]{.95\textwidth}
  \includegraphics[width=8cm]{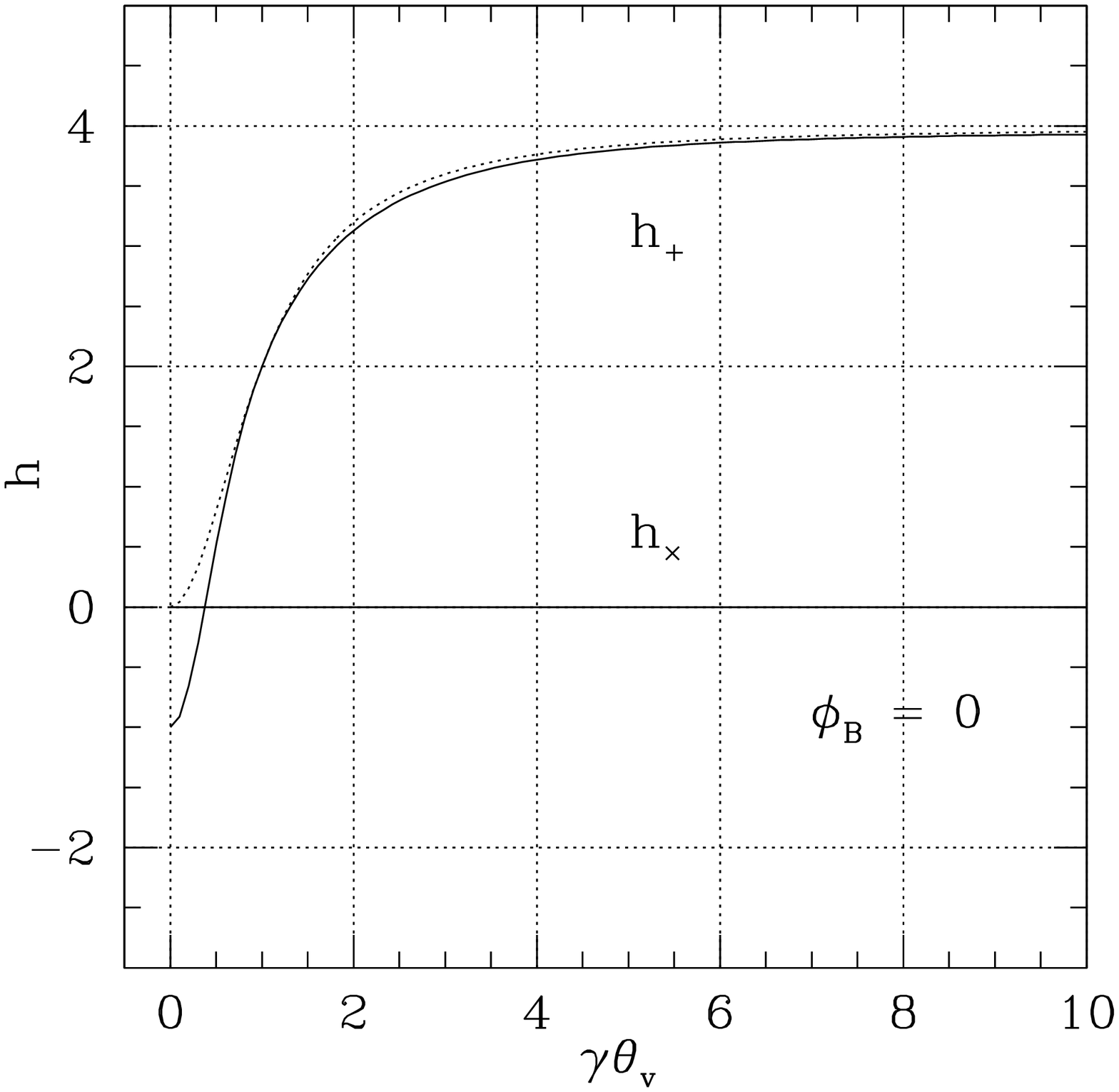}
 \hskip 0cm
  \includegraphics[width=8cm]{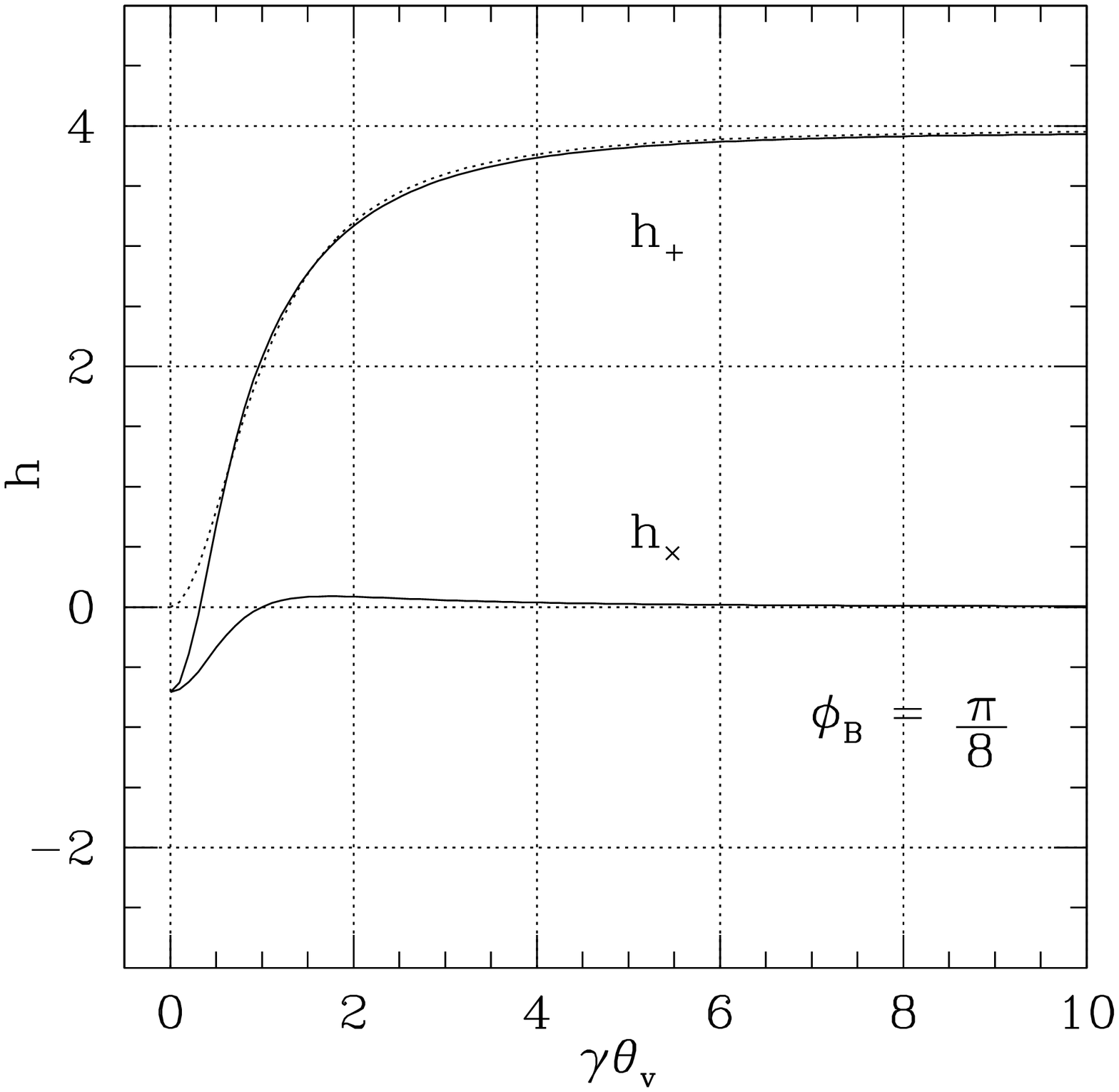}
 \end{minipage}
\vskip 0cm
\hskip 0cm
\begin{minipage}[t]{.95\textwidth}
   \includegraphics[width=8cm]{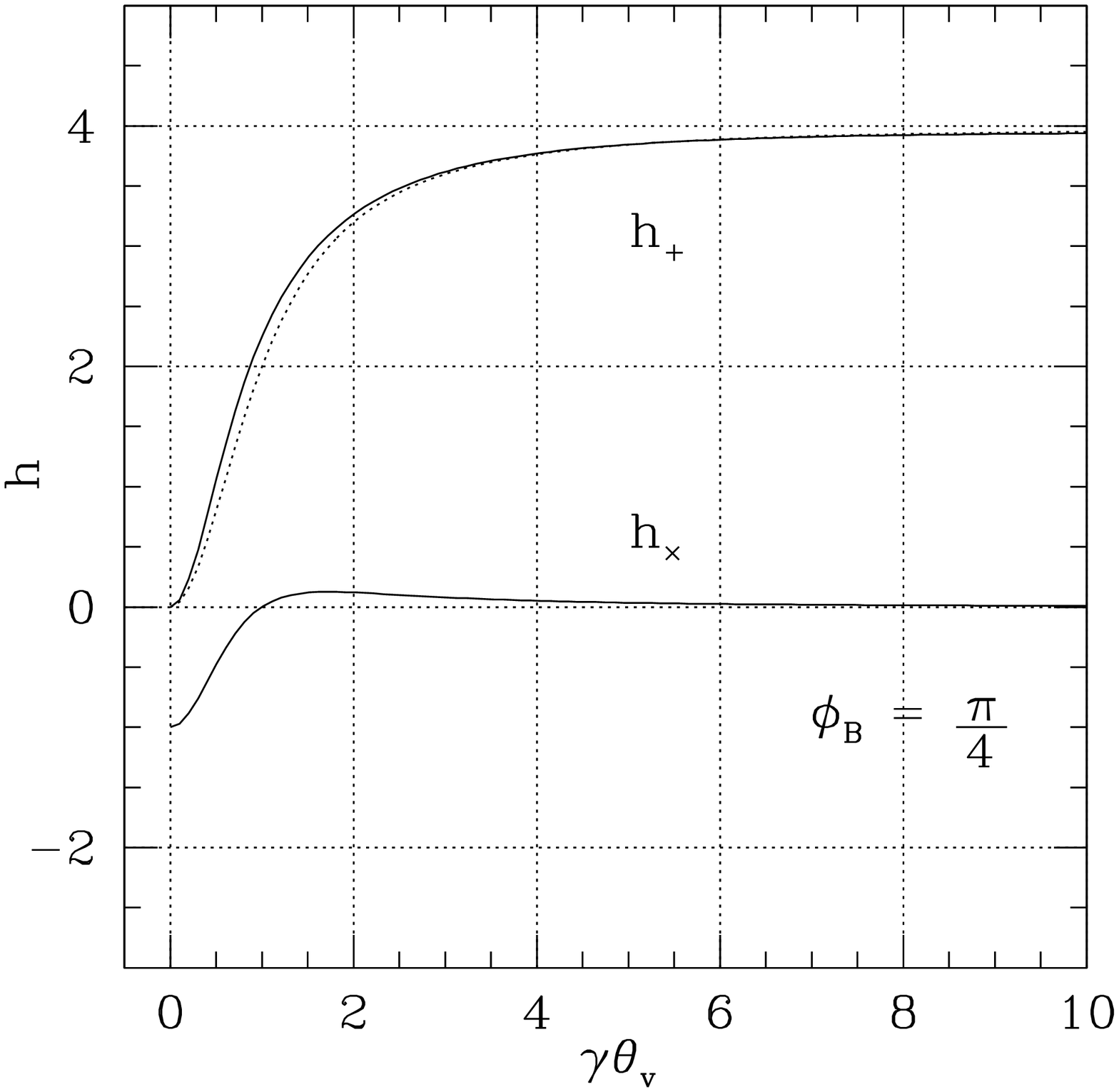}
 \hskip 0cm
  \includegraphics[width=8cm]{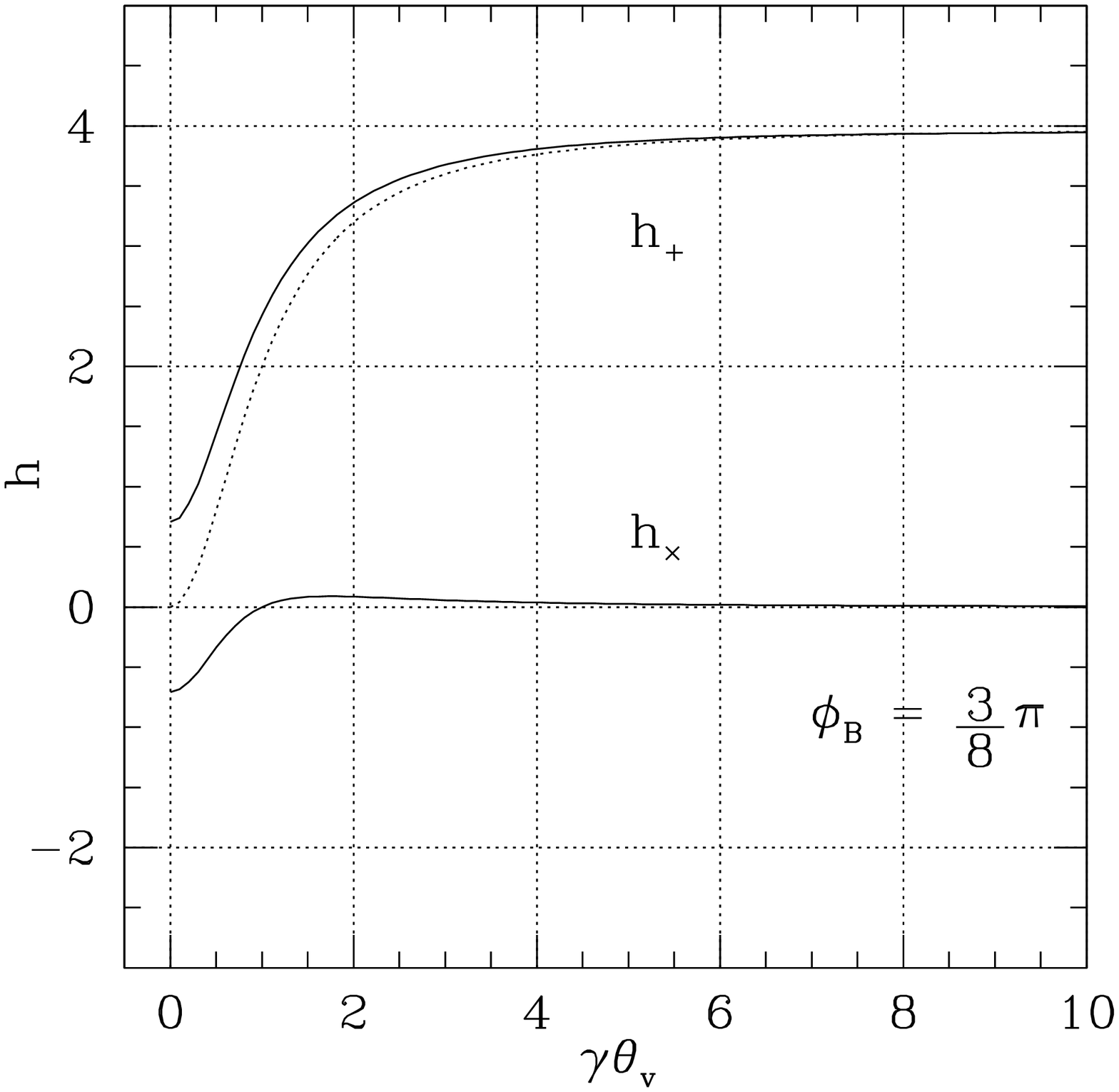}
 \end{minipage}
 \caption{
GW memory from photons with anisotropic angular distribution of equation (19)
normalized by \(E/d\) (case of synchrotron emission with ordered magnetic fields).
Two polarization components \(h_+\) and \(h_{\times}\) (solid lines) are shown for 
\(\phi_0 = 0\) (top left), \(\frac{\pi}{8}\) (top right), 
\(\frac{\pi}{4}\) (bottom left), and \(\frac{3}{8} \pi\) (bottom right)
as a function of \(\gamma \theta_v \).
The memory from a point mass with the same energy \(\gamma m = E\) is shown for comparison (dotted line).}
\end{figure*}

\begin{figure*}
\begin{minipage}[t]{.95\textwidth}
  \includegraphics[width=8cm]{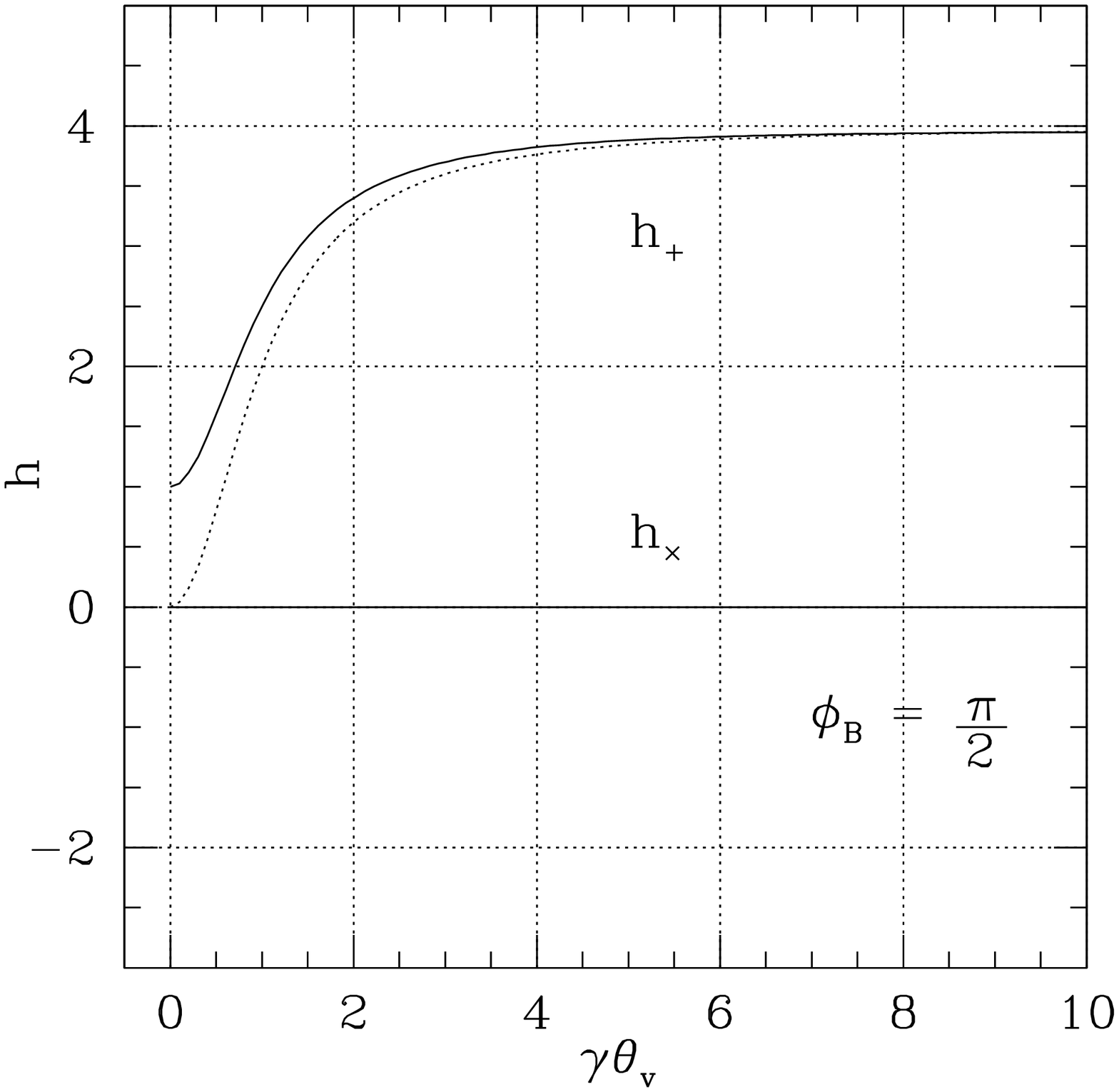}
 \hskip 0cm
  \includegraphics[width=8cm]{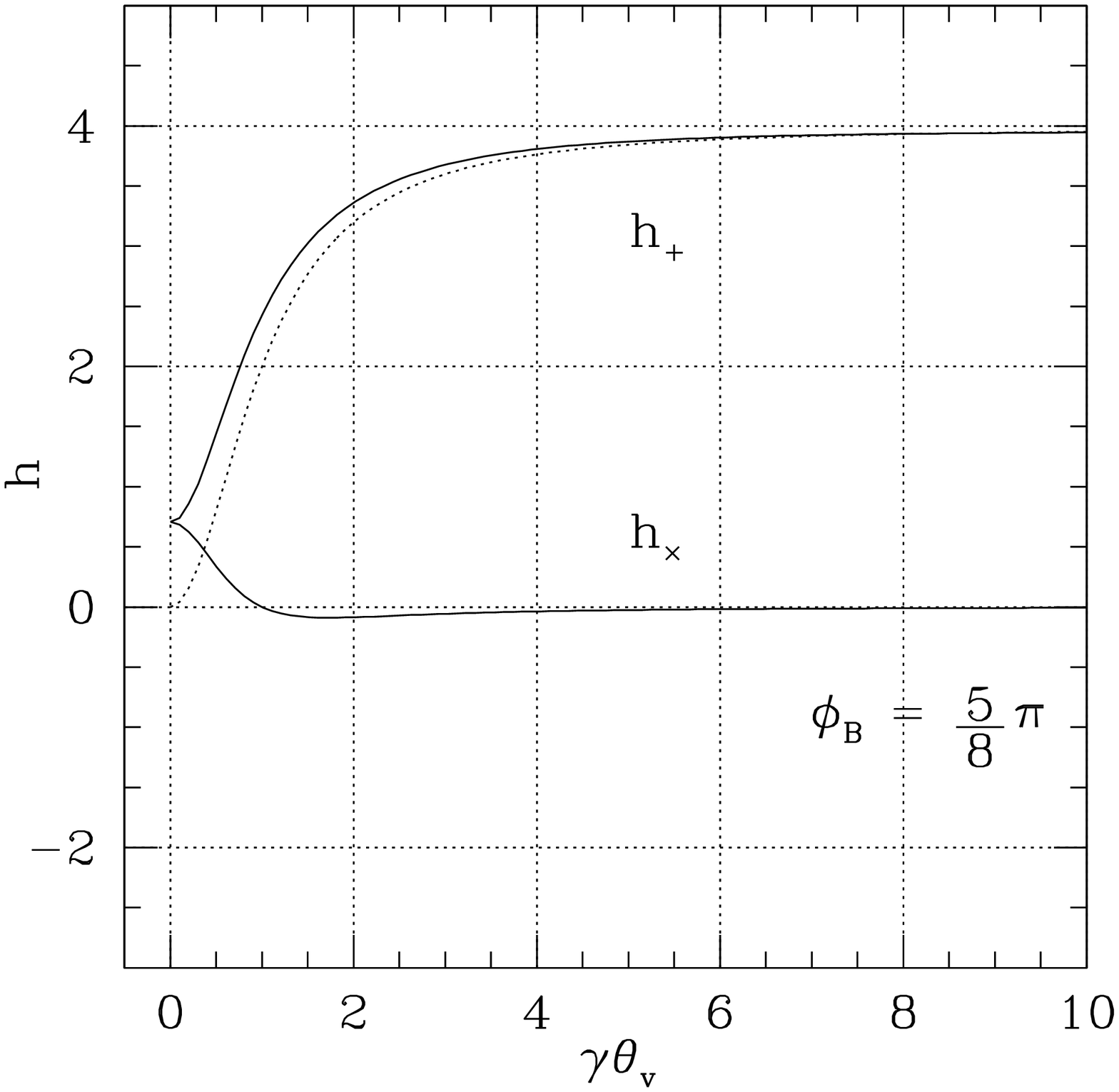}
 \end{minipage}
\vskip 0cm
\hskip 0cm
\begin{minipage}[t]{.95\textwidth}
   \includegraphics[width=8cm]{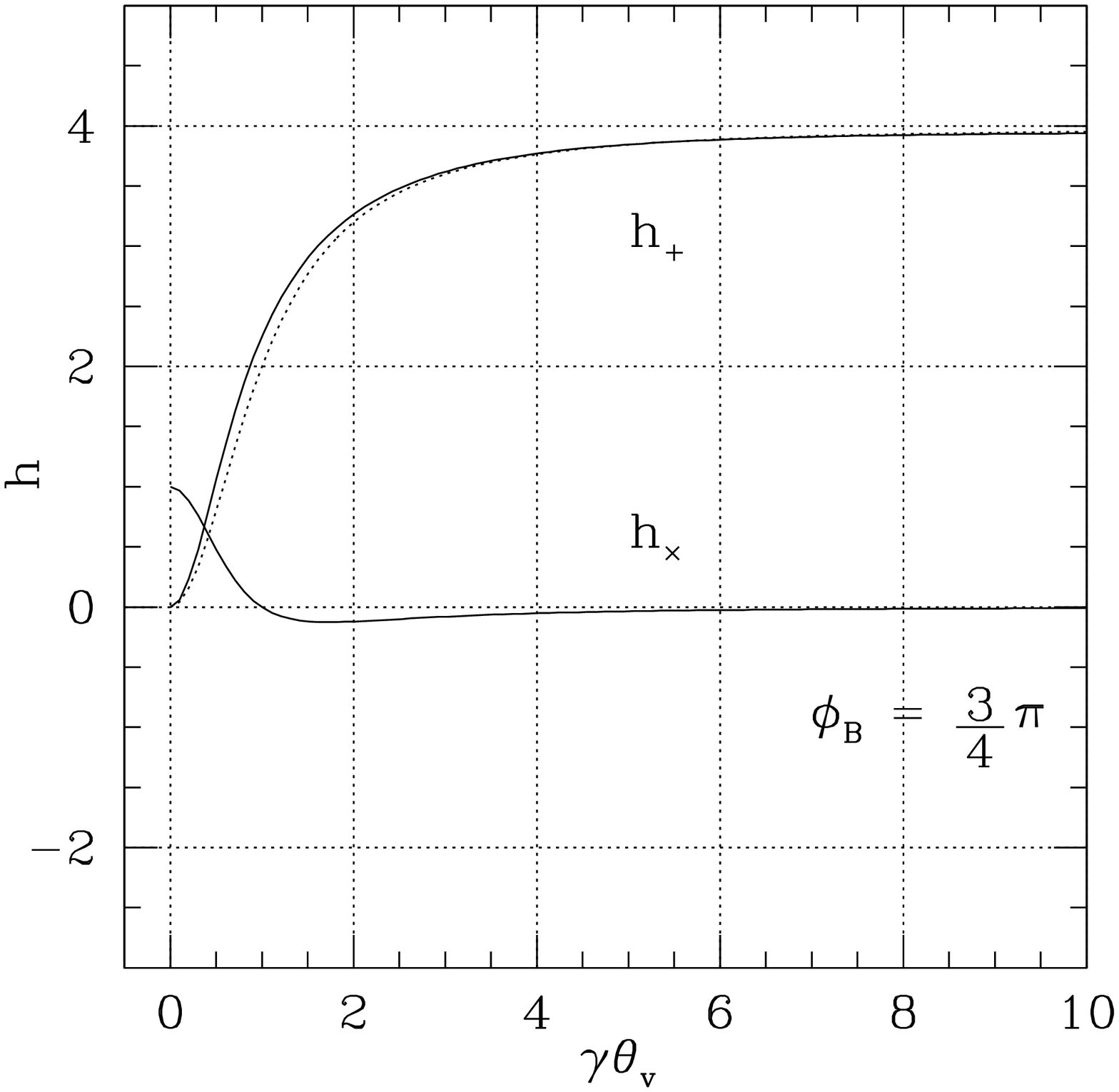}
 \hskip 0cm
  \includegraphics[width=8cm]{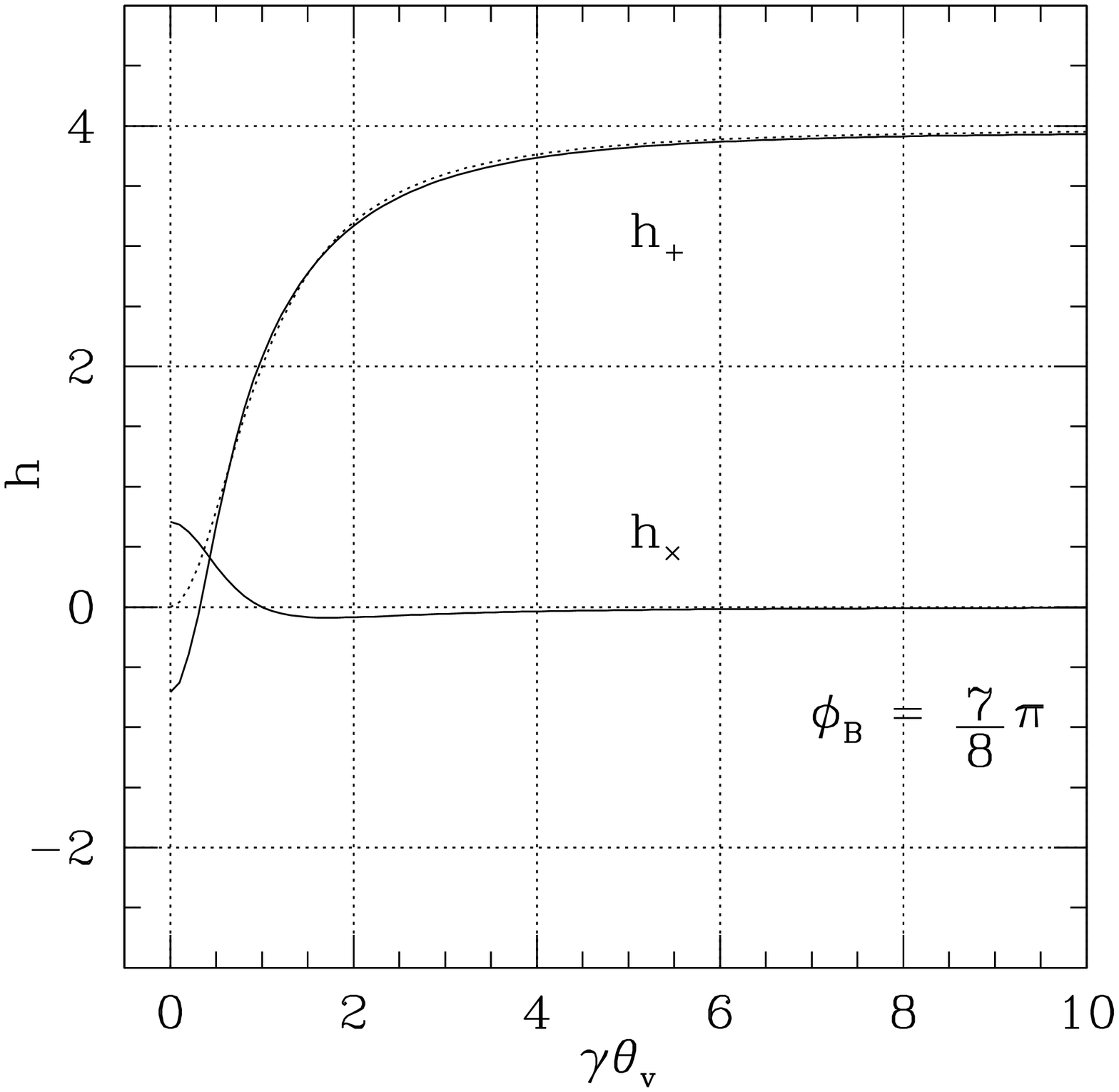}
 \end{minipage}
  \caption{
The same with Fig.7 but 
for \(\phi_B = \frac{\pi}{2} \) (top left), \(\frac{5}{8} \pi\) (top right), 
\(\frac{3}{4} \pi \) (bottom left), and \(\frac{7}{8} \pi\) (bottom right).
GW memory from photons with anisotropic angular distribution of equation (19)
normalized by \(E/d\)  (case of synchrotron emission with ordered magnetic fields).
The amplitude from a point mass with the same energy \(\gamma m = E\) is shown for comparison (dotted line).}
\end{figure*}

Now we consider the variation in the GW memory as a result of photon emission from 
a point mass. If a point mass \(m\) with a Lorentz factor \(\gamma\) ( velocity \(\beta ) \) radiates photons with power \(P\) 
during a short time interval \(\Delta t\), the change in the total GW memory
\(\Delta h = \Delta h_+ + i \Delta h_{\times} \) is given by

\begin{eqnarray}
\Delta h &=& \frac{2 \gamma' m}{d} \frac{\beta'^2 \sin^2 \theta_v}{1-\beta' \cos \theta_v}
+ \frac{P \Delta t}{d} < 2(1+\cos \theta) e^{2 i \phi} > \nonumber \\
&& -\frac{2 \gamma m}{d} \frac{\beta^2 \sin^2 \theta_v}{1-\beta \cos \theta_v},
\end{eqnarray}

\noindent
where \(\gamma' (\beta')\) is the Lorentz factor (velocity) of the point mass after time \(\Delta t \).
Since the conservation of energy implies \(\gamma' m = \gamma m - P \Delta t\), we find
that the following relation holds to first order in \(\Delta t\), 

\begin{eqnarray}
\Delta h &=& \frac{2 \gamma m}{d} \Delta \left( \frac{ \beta^2 \sin^2 \theta_v}{1-\beta \cos \theta_v}
\right) \nonumber \\
&&+ \frac{P \Delta t}{d} \left( 
 < 2(1+\cos \theta) e^{2 i \phi} > 
-\frac{2 \beta'^2 \sin^2 \theta_v}{1-\beta' \cos \theta_v} 
\right) \nonumber \\
&=& -\frac{2P \Delta t}{d} \frac{(2-\cos \theta_v) \sin^2 \theta_v }{\gamma^2 (1-\beta \cos \theta_v)^2} \nonumber \\
&&+ \frac{P \Delta t}{d} \left( 
 < 2(1+\cos \theta) e^{2 i \phi} > 
-\frac{2 \sin^2 \theta_v}{1-\beta \cos \theta_v} 
\right),
\end{eqnarray}

\noindent
where we assume \(\gamma, \gamma' >> 1 \) and retain the terms in leading orders of \(\gamma^{-2}\).
In equation (22) the first term is the variation in the memory of the point mass due to broadening of
the hole of anti-beaming, while the second term is the change in the memory ascribed to the  emission of photons. The size of the terms in the parenthesis of equation (22) can be easily read from figures 3, 4, 5, 7, and 8 as the difference between \(h_+\) (solid line) and the memory from a point mass (dotted line) as well as
 the value of \(h_{\times}\). For the case of isotropic emission, the second term is negligibly small compared to the first term ( Figure 3). Thus the memory decreases monotonically as the point mass slows down, even if
the contribution from photons is added. However, for the case of anisotropic emission 
above considered (equations 15 and 19), the second term in equation (22) makes a 
dominant contribution
to the change in the memory compared to the first term for small \(\gamma \theta_v\).
For large \(\gamma \theta_v\) the first term in equation (22) is smaller than the second term in the parenthesis
by a factor of \(\gamma^2\) and thus it is negligible again.
As a consequence, for \(\gamma \theta_v < 1\)  and 
even for 
\(\theta_v = 0\) the GW memory varies by a significant fraction of its maximum value  
on the conversion of energy from a point mass into photons.

\section{Gravitational Wave Memory from the Decelerating Phase of GRB Jets}

Here we estimate the overall behavior of time variation in the GW memory from GRB jets 
by applying our results in \S 2 to specific models of GRB phenomena.
It has been shown that the temporal structure of GRBs are naturally explained by the internal shock model, 
in which gamma-rays are radiated in an inhomogeneous relativistic wind possibly generated by a variable central engine (\cite{Piran1999}). 
Such a wind has been modeled by multiple shells moving in procession with various Lorentz factors.
It is assumed that if a rapid shell catches up a slower shell ahead the two shells would collide and merge converting a fraction of kinetic energy into thermal energy. The thermal energy released will be 
radiated on the spot. We follow the formulation by \citet{Kobayashi1997} to model 
the internal shock.
Here we use the internal shock model only to reproduce the time variability of typical GRB light curves.

We assume that a rapid shell with mass \(m_r\) and Lorentz factor \(\gamma_r\) collides with 
a slow shell with mass \(m_s\) and Lorentz factor \(\gamma_s\) to form a merged shell with mass
\(m_r + m_s\) and Lorentz factor \(\gamma_m\). The conservation of energy and momentum leads to

\begin{eqnarray}
m_r \gamma_r + m_s \gamma_s &=& (m_r + m_s + \epsilon) \gamma_m, \nonumber \\
m_r \sqrt{\gamma_r^2-1} + m_s \sqrt{\gamma_s^2-1} 
&=& (m_r + m_s + \epsilon) \sqrt{\gamma_m^2-1}, 
\end{eqnarray}

\noindent
where \(\epsilon\) is the internal energy being released
in the rest frame of the merged shell. 
For large \(\gamma_r, \gamma_s >> 1\), \(\gamma_m\) is approximately given by

\begin{equation}
\gamma_m \simeq \sqrt{\frac{m_r \gamma_r + m_s \gamma_s}{m_r/\gamma_r + m_s/\gamma_s}},
\end{equation}
 
\noindent
and then the energy converted into radiation \(E\) is estimated to be 

\begin{eqnarray}
E = \gamma_m \epsilon = m_r (\gamma_r -\gamma_m) + m_s (\gamma_s-\gamma_m).
\end{eqnarray}

At the time of collision the forward and reverse shocks arise. The Lorentz factors of the forward 
and reverse shocks, \(\gamma_{fs}\) and \(\gamma_{rs}\), are given by 

\begin{eqnarray}
\gamma_{fs} \simeq \gamma_m \sqrt{ \left( 1+ \frac{2 \gamma_m}{\gamma_s} \right) /
\left( 2 + \frac{\gamma_m}{\gamma_s} \right) }, \nonumber \\
\gamma_{rs} \simeq \gamma_m \sqrt{ \left( 1+ \frac{2 \gamma_m}{\gamma_r} \right) /
\left( 2 + \frac{\gamma_m}{\gamma_r} \right) }, \nonumber 
\end{eqnarray}

\noindent
respectively (\cite{Sari1995}). The time in which gamma-rays are emitted is approximately
given by the time in which the reverse shock traverses the rapid shell,

\begin{equation}
\Delta t = \frac{l_r}{\beta_r-\beta_{rs}},
\end{equation}

\noindent
where \(l_r\) is the width of the rapid shell and the \(\beta_{rs}\) is the velocity of the
reverse shock. We note that the emitting region moves at a speed of \(\beta_m\) and that 
the observed timescale is reduced by a factor of
\(1-\beta_m \cos \theta_v\), which is \(\sim 1/2 \gamma_m^2\) for \(\theta_v = 0\). 

The width of the merged shell \(l_m\) is simply given by widths of the rapid and slow shells
\(l_r, l_s\) as

\begin{eqnarray}
l_m = l_s \frac{\beta_{fs}-\beta_m}{\beta_{fs}-\beta_s}
+ l_r \frac{\beta_m-\beta_{rs}}{\beta_r-\beta_{rs}}.
\end{eqnarray}

In calculating the GW memory, we assume that for each collision the energy of 
photons \(E\) (equation 25) is radiated in time \(\Delta t\) (equation 26) so that the emission 
power is given by \(P = E/\Delta t\). Applying equation (21) in the last section, the change 
in the GW memory for a single 
collision is evaluated as 

\begin{eqnarray}
\Delta h(\theta_v, \phi_B) = \Delta h_{{\rm photons}}(\theta_v, \phi_B)
+ \Delta h_{{\rm jet}}(\theta_v),
\end{eqnarray}

\noindent
where 

\begin{eqnarray}
&& \Delta h_{{\rm photons}}(\theta_v, \phi_B) \nonumber \\
&&
 = \frac{m_r (\gamma_r-\gamma_m)+m_s (\gamma_s -\gamma_m)}{d} 
 <2 (1+\cos \theta) e^{2i \phi}>
\end{eqnarray}

\noindent
for photons' contribution and 

\begin{eqnarray}
\Delta h_{{\rm jet}}(\theta_v) =& & \frac{2 m_r \sin^2 \theta_v}{d} \left( 
\frac{\gamma_m \beta_m^2}{1-\beta_m \cos \theta_v}
- \frac{\gamma_r \beta_r^2}{1-\beta_r \cos \theta_v} \right) \nonumber \\
 +
\frac{2 m_s \sin^2 \theta_v}{d}&& \left( \frac{\gamma_m \beta_m^2}{1-\beta_m \cos \theta_v}
- \frac{\gamma_s \beta_s^2}{1-\beta_s \cos \theta_v} \right)
\end{eqnarray}

\noindent
for the fluid contribution in the shells. We assume that this change in the memory occurs during time \(\Delta t\)
at a constant rate. We study the case of synchrotron emission with ordered magnetic fields for \(\Delta h_{{\rm photons}}\). 
We always have a real and negative \(\Delta h_{{\rm jet}}\), while \(\Delta h_{{\rm photons}}\) may be
complex reflecting the non-axisymmetric distribution of the photon emission.

Now we describe the method for simulating the collision of the multiple shells in the internal shock. 
We take \(N\) shells with the same mass \(m\) and width \(l\)
(i.e., the same density) being placed with an equal separation \(L\) at time \(t=0\). 
To each shell a Lorentz factor, uniformly distributed
between \(\gamma_{{\rm min}}\) and \(\gamma_{{\rm max}}\), is assigned at the outset. 
Given the Lorentz factor \(\gamma_i\), the velocity \(\beta_i\), and the position
\(x_i = -(L+l) i \) for \(i\)-th shell
( \(i = 1, \cdots, N\) ) as the initial conditions, we follow the time evolution of the position
 of all the shells until the first collision occurs. At the collision two shells are merged and again the
movement of the shells is followed till the next collision. The same procedure is repeated
until either all shells will be merged into a single large shell 
or we get the configuration where the shells are marching
with decreasing velocities from the head to the end. We used parameters 
\(N = 100\), \(\gamma_{{\rm min}} = 10^2, 
\gamma_{{\rm max}} = 10^3\), and \(L=1, l=0.1\). The overall evolution of the GW memory 
with time and the gamma-ray light curve are determined by the ratio 
\(L/l\), \(N\), \(\gamma_{{\rm min}}\), \(\gamma_{{\rm max}}\) and \(\theta_v\).

Since the GW memory is polarized 
in the direction from the source to the line of sight on the transverse plane,   
we need to account for the finiteness of the conical opening half-angle of the
GRB jets to estimate the net memory.
The typical size of the opening half-angle has been estimated to be 
\(\Delta \theta \sim 0.1\) rad, which is much larger than the beaming angle \(\gamma^{-1}
< 0.01 \) (\cite{Frail2001}).
If we consider the memory from an axisymmetric jet that is seen nearly head-on, 
not only the memory from the central part of the jet vanishes but also the memory added up from 
the edge parts turns out to be canceled because of the symmetric distribution of polarization angles.
Then the GW memory from the acceleration phase of the GRB jets 
can be observed only if the jet is seen off-axis \(\theta_v > \Delta \theta\) (\cite{Sago2004}).   
We note that this conclusion depends on the geometry of the jet. If the jet is uniform 
within its opening angles, the GW memory from the jet still has about half of its maximum values
at \(\theta_v = \Delta \theta\).

Similar effects of the finite opening half-angle exist 
for the change in the GW memory from the decelerating phase of the GRB jets.   
For isotropic gamma-ray emission in the rest frame of the jet, the GW memory
from photons is found to be nearly equal with that from the  point mass that emits the photons  
( \S 2 ). Thus, the memory decreases monotonically owing to the decline in the Lorentz factor
and the resultant expansion of the 
anti-beaming hole.  In this case the GW memory might be observable
only if \(\theta_v > \Delta \theta\) as in the case of the acceleration phase. 

\begin{figure}
  \begin{center}
\includegraphics[width=8cm]{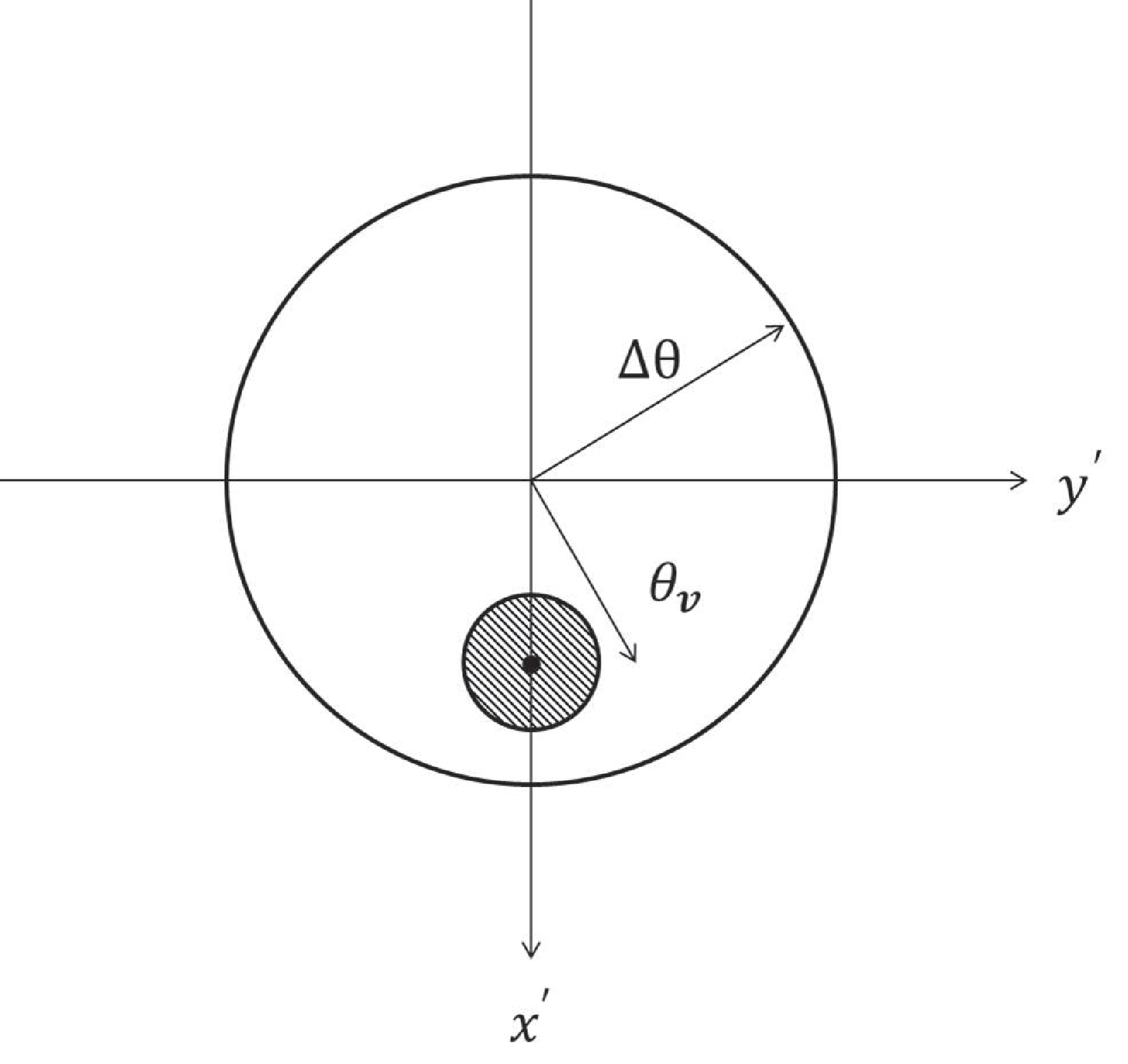}
    %%% \FigureFile(width,height){filename}
  \end{center}
  \caption{A uniform jet with an opening half angle \(\Delta \theta\) is seen 
from the top of the jet. Only a small portion 
of the jet around the line of sight of an angular size \(\gamma^{-1}\) (shaded circle) contributes to the change in the GW memory.}
\end{figure}

Recent successes in the detection of the polarization in the prompt emission of GRBs
(\cite{Yonetoku2011})
suggests the presence of ordered magnetic fields and/or an anisotropic photosphere in the emission region. 
The coherence length scale of the magnetic fields is not well understood 
for any GRBs and the origin of such ordered magnetic fields is still a debated issue. 
A variety of different magnetic field configurations and viewing angles have been
considered to explain a large degree of polarization (\cite{Granot2003}; \cite{Lazzati2006}; \cite{Toma2009}).  
If the angular scale of coherent magnetic field is small compared to the beaming angle 
\(\gamma^{-1}\) or the magnetic fields are randomly oriented, there is no preferred azimuthal angles
into which the gamma-ray emission is focused so that the emission ought to be essentially isotropic. 
Then the change in the GW memory on each collision should be quite small
so that we would not observe a variation in the memory from the decelerating phases of GRB jets.

Instead if the shell contains a number of patches of locally ordered magnetic fields
we would have a non-vanishing variation in the memory for some viewing angles.
The case in which we would expect the largest change in GW memory is 
when an ordered transverse magnetic field exists in the relativistic shell.
We consider a uniform jet with an opening half angle  
\(\Delta \theta\) illustrated in Figure 9. The line of sight ( \(z\) axis ) is located at a point tilted from \(z'\) axis 
by an angle \(\theta_v\)
toward \(x'\) direction.
We assume that the magnetic field has only a transverse component and that
the field line is parallel to the meridional plane at \(\phi' = \alpha\). Then the unit vector
\(\hat{\mathbf{B}} = \mathbf{B}/|\mathbf{B}| \) at a point \( (\theta', \phi')\) is expressed as

\begin{eqnarray}
\hat{\mathbf{B}} &=& 
\frac{1}{\sqrt{\sin^2 \theta' \cos^2 (\phi'-\alpha) + \cos^2 \theta'}}
\nonumber \\
&& \left( \cos \theta' \cos \alpha, 
\cos \theta' \sin \alpha,
-\sin \theta' \cos (\phi'-\alpha)
\right).
\end{eqnarray}

\noindent
We study the case of \(\alpha = \pi/2\), in which the highest degree of anisotropy is expected for GW memory.   
In order to use \(\Delta h(\xi, \phi_B)\) calculated for the synchrotron model (Equation 28), we define \(\phi_B\) such that
\(\cos \phi_B = \hat{\mathbf{B}} \cdot \hat{\mathbf{t}}\) where \(\hat{\mathbf{t}}\) is
a unit vector that is tangent to the great circle connecting
the points \( (\theta', \phi')\) and \((\theta_v, 0)\)
at \((\theta',\phi')\),
 
\begin{eqnarray}
\hat{\mathbf{t}} &=& \frac{\hat{\mathbf{p}}-\hat{\mathbf{q}} \cos \xi}{\sin \xi}, 
\end{eqnarray}

\noindent
where 
\begin{eqnarray}
\hat{\mathbf{p}} &=& ( \sin \theta_v, 0, \cos \theta_v ), \nonumber \\
\hat{\mathbf{q}} &=& ( \sin \theta' \cos \phi', \sin \theta' \sin \phi', \cos \theta' ),
\end{eqnarray}

\noindent
and \(\cos \xi = \hat{\mathbf{p}} \cdot \hat{\mathbf{q}}\). The angle \(\xi\) is
equal to \(\theta\), the polar angle in \(xyz\) frame.
The GW memory is calculated by summing up additive contributions to the amplitude from all points in the shell as follows.

\begin{eqnarray}
h = \int \frac{\sin \theta' d \theta' d \phi'}{\Delta \Omega}
\Delta h(\xi,\phi_B) e^{2 i \phi}, 
\end{eqnarray}

\noindent
where \(\Delta \Omega = 2 \pi (1-\cos \Delta \theta)\) and the integration is 
performed over the angles inside the jet.
We note that equation (12) relates \((\theta, \phi)\) to \((\theta',\phi')\).

\begin{figure*}
\hskip 0.3cm
\begin{minipage}[t]{.95\textwidth}
  \includegraphics[width=8cm]{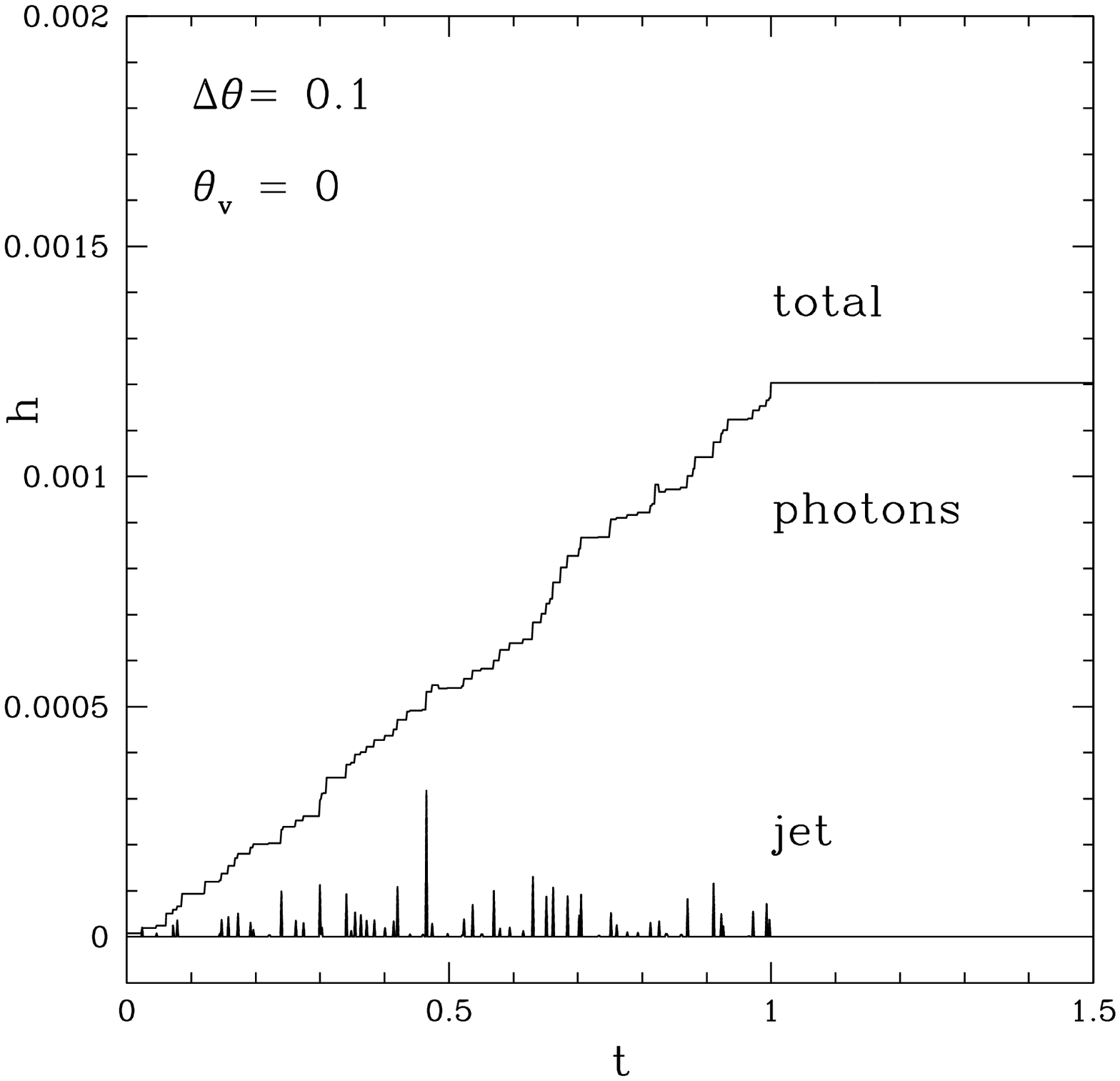}
 \hskip 1cm
  \includegraphics[width=8cm]{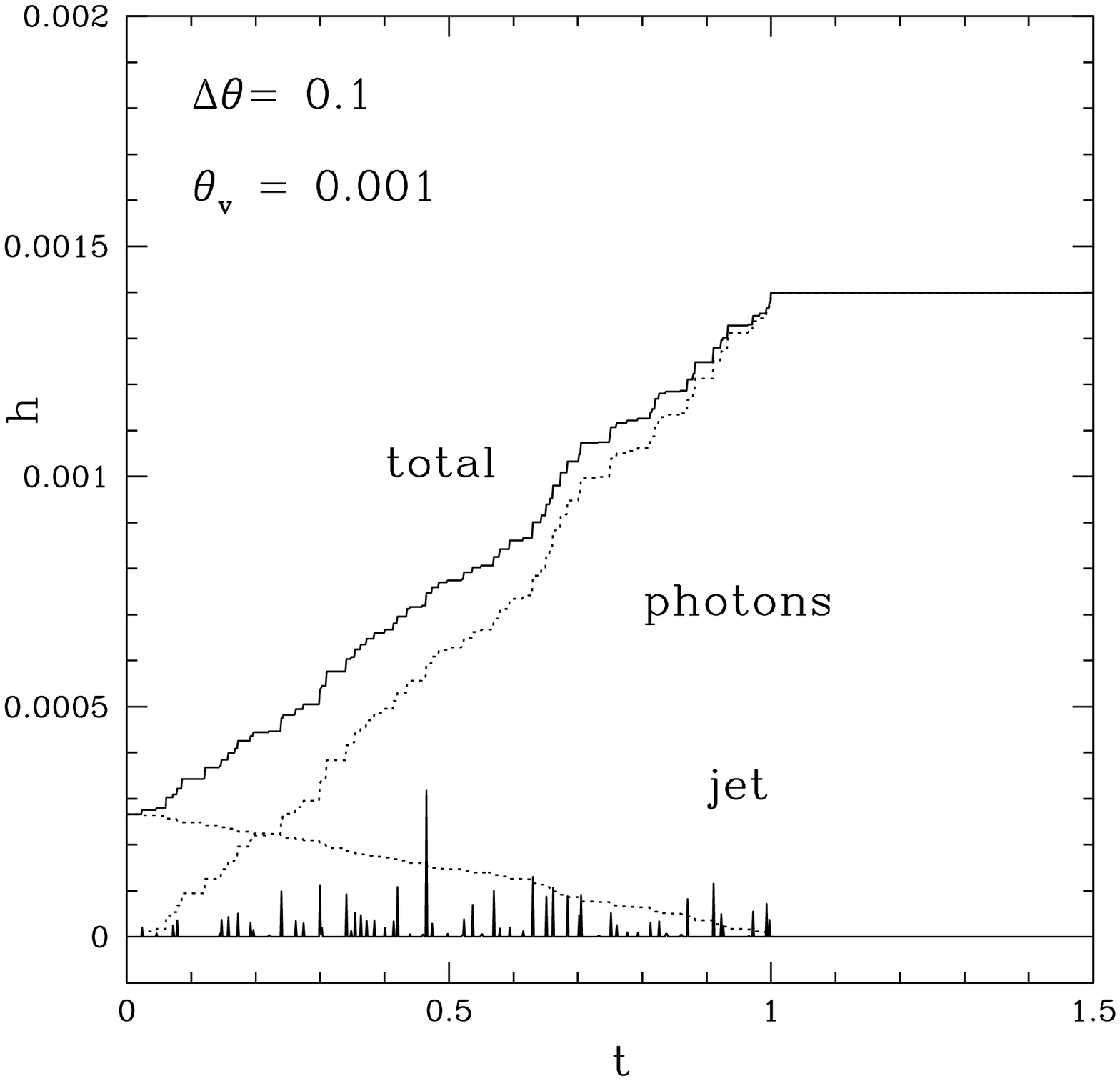}
 \end{minipage}
  \caption{The amplitudes of 
GW memory \(h = \sqrt{|h_+|^2 + |h_{\times}|^2}\) normalized by \(E_{\gamma}/d\), where \(E_{\gamma}\) is 
the total energy radiated in gamma-rays, are shown for the case of synchrotron emission with ordered magnetic
fields with \( \Delta \theta = 0.1\) and \(\alpha = \pi/2\).    
The two panels are GW memory for different viewing angles
\( \theta_v = 0\) (left) and \(\theta_v = 0.001\) (right). 
The time \(t\) is arbitrarily scaled.
The gamma-ray pulses, also plotted with an arbitrary scale, correspond to the collision events, which 
are modeled by  internal shocks with \(N = 100 \) relativistic shells of Lorentz factors 
\(\gamma = 10^2 - 10^3\).}
\end{figure*}

Figure 10 shows the GW memory \(h = \sqrt{|h_+|^2 + |h_{\times}|^2}\)
calculated for the case of synchrotron emission with ordered magnetic fields 
with \(\Delta \theta = 0.1\) and \(\alpha = \pi/2\). We show two plots for cases with different viewing angles, 
\( \theta_v = 0\) (left) and \(\theta_v = 0.001\) (right). 
The time is arbitrarily scaled.
The time which passed till the memory reaches the plateau corresponds to the duration of the gamma-ray emission of the GRB.
The amplitude is normalized by \(E_{\gamma}/d\), where \(E_{\gamma}\) is the total energy radiated in 
gamma-rays. The value of \(h_{{\rm jet}}\) is shifted so that the final level of \(h_{{\rm jet}}\) turns out to be zero.  
The gamma-ray pulses, also plotted with an arbitrary scale, correspond to the collision events. 
 We see that the memory from photons emitted at each collision is added up, leading to
a monotonic rise in net memory with time.
For \(\theta_v = 0 \) the memory from the jet is suppressed and unobservable throughout.
We ignored the broadening of the gamma-ray light curves that should appear because of  
the finite opening angle of the jet. 
The rise of the
GW memory at each collision should also be smeared. However, even if such smearing is taken into account, the waveform, which  
is made from cumulative contributions, would not change its shape so much. 

The behavior of gravitational waveforms shown in Figure 10 can be understood as follows.  As we changed terms in equation (21) as equation (22), we rearrange terms in
equations (28), (29), and (30).  Then, by leaving only dominant terms,
we have 

\begin{eqnarray}
\Delta h(\theta_v, \phi_B)
\sim \frac{E}{d} \left( 
<2(1+\cos \theta) e^{2i\phi}>
- \frac{2 \sin^2 \theta_v}{1-\beta_m \cos \theta_v} 
\right),
\end{eqnarray}

\noindent
where \(E = m_r (\gamma_r-\gamma_m)+m_s (\gamma_s -\gamma_m)\).

\noindent
As is found in \S 2, \(\Delta h(\theta_v,\phi_B)\) takes a non-zero value
of order \(\sim E/d\) only for \(\gamma \theta_v\)  less than \(\sim 1 \) 
 (Figures 7 and 8). 
Thus, we use equation (35) in (34) and change variables of integration 
into \(\theta, \phi\) to obtain
a rough estimate

\begin{eqnarray}
h \sim \frac{E_{\gamma}}{d} ( \gamma \Delta \theta )^{-2},
\end{eqnarray}

\noindent
where \(E_{\gamma}\) is the energy radiated as photons. This is valid both
for a single collision and for the sum of radiated energy for all the collisions.
For \(\Delta \theta = 0.1\) and \(\gamma \sim 3 \times 10^2\) we
have \(h \sim 10^{-3} E_{\gamma}/d \). This is a change of the GW memory.
The initial amplitude of the GW memory is carried by the uniform jet and its approximate
value is given by

\begin{eqnarray}
h_0 \sim \frac{4E_{{\rm jet}}}{d} \frac{\theta_v^2}{\Delta \theta^2+
\theta_v^2},  
\end{eqnarray}

\noindent
where \(E_{\rm  jet}\) is the kinetic energy of the relativistic jet.
We have confirmed that this approximate expression agrees with the exact value
(\cite{Sago2004}; \cite{Hiramatsu2005}) to an accuracy of 10 percent
for \(\gamma= 10^2 - 10^3\) and a wide range of \(\Delta \theta\). If we assume \(E_{\rm jet} = E_{\gamma} \), 
we obtain
\(h_0 \sim 4 \times 10^{-4} E_{\gamma}/d \) for \(\Delta \theta = 0.1\) and
\(\theta_v = 0.001\),  being in a qualitatively good agreement with the results
shown in Figure 10.

As is seen in Figures 7 and 8, the change in the GW memory is seen only within 
small angles of \(\gamma^{-1}\) from the line of sight. 
As long as the small solid angles around the line of
sight ( a shaded circle in Figure 9 ) is included in the jet with an opening half angle \(\Delta \theta\), which is equivalent to \(\theta_v < \Delta \theta -1/\gamma\),
we would observe almost the same variation in the GW memory regardless of the position of the line of sight. This implies that an ordered magnetic field is required only within 
a small solid angle from which gamma-rays we observe are radiated.

\section{Discussion and Summary}

We studied the GW memory from a radiating and decelerating point mass.
We calculated the memory from the photons averaged over angular distribution of the emission.
For isotropic emission (in the rest frame of the point mass), the
averaged memory has quite similar dependence on the viewing angle as the memory from the point mass. 
That is, the memory from the photons is strongly suppressed for small viewing angles.
However if the photons are emitted anisotropically, the averaged GW memory 
may have a large amplitude even for small viewing angles, enabling us to
observe the GW memory of GRB jets. 
We demonstrated an example of the gravitational waveform expected
for the internal shock model in the case of synchrotron emission with an ordered magnetic field. We find that the GW memory shows a continuous rise
over the time scale of the gamma-ray emission as well as tiny jumps 
correlated with gamma-ray pulses.  
Such an anisotropic emission of gamma-rays has been inferred from the detection of polarizations
in the GRB emissions (\cite{Steele2009}; \cite{Yonetoku2011}). 
Obviously, we need more samples of such polarization observations
and further understanding of the relativistic jet's configuration and the mechanism 
of gamma-ray emission to consider realistic modeling of the GW memory
from GRB jets. 

\begin{figure}
\begin{center}
\hskip 0cm
  \includegraphics[width=8cm]{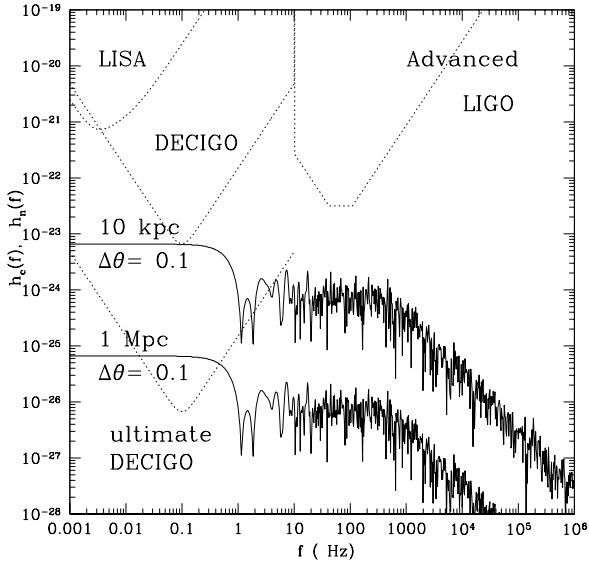}  
\end{center}
  \caption{The characteristic amplitude \(h_c(f)\) of the GW memory from the  decelerating phase
of a GRB jet with \(E_{\gamma} = 10^{51}\) erg. 
Synchrotron emission from electrons in ordered magnetic fields is assumed.
Solid lines represent \(h_c(f)\) for two cases 
of \((d, \Delta \theta) = (1 {\rm Mpc}, 0.1), (10 {\rm kpc}, 0.1)\). 
We assume \( \alpha = \pi/2, \theta_v = 0\) for both cases. The amplitude scales as \(\Delta \theta^{-2}\), 
since we fix \(E_{\gamma}\) and only a small portion of the jet of angular size \(\sim \gamma^{-1}\)
contributes to the change in the GW memory.
The amplitude does not depend
on \(\theta_v\) as long as \(\theta_v\) is relatively small compared to \(\Delta \theta\).
We have not shown cases with \(\theta_v > \Delta \theta \), since we would then have a significantly small variation
 in the GW memory so that such cases are out of interest.
 }
\end{figure}
The GW memory would also provide clues to discriminating
different models of GRBs and related GRB phenomena.
Based on the unified model of GRBs the relativistic jet is composed of
a number of sub-jets launched from a central engine (\cite{Yamazaki2004}).  
In this case each sub-jet would be threaded with ordered magnetic fields 
that might be oriented in various directions. 
Then the polarization angles of the GW memory from each sub-jet 
may be different, which leads to a sharp fluctuation of the memory with a relatively small amplitude. 
The unified model also predicts that relativistic jets seen with large viewing angles
are observed as XRFs. For such cases with moderate viewing angles the GW 
memory should vary owing to the broadening of the anti-beaming hole. 
Whether the memory has such features in its time variation could help us to test GRB models.

The GW memory from decelerating GRB jets is expected to have characteristic frequencies
\(0.1 - 10 \) Hz,
corresponding to the duration of GRBs, \(T \sim 0.1 - 10\) sec, so that it is a suitable target
for DECIGO and BBO (\cite{Seto2001}; \cite{Sago2004}) . 
Figure 11 shows the characteristic amplitude \(h_c(f) =
f |\tilde{h}(f)| \), where \(\displaystyle{ 2 \pi i f \tilde{h}(f) = \int_{-\infty}^{\infty} \dot{h}(t) e^{- 2 \pi i f t} dt }\),
from the decelerating phase of a GRB jet with \(E_{\gamma} = 10^{51}\) erg and a duration of \(1\) sec. 
The noise amplitude \(h_n(f) = [ 5f S_h(f)]^{1/2}\), where \(S_h(f)\) is the spectral density of the
strain noise in the detector at frequency \(f\), is also shown for Advanced LIGO, LISA, and DECIGO/BBO. 
We use the same formula for \(h_n(f)\) or \(S_h(f)\) with the one used in \citet{Sago2004} and \citet{Suwa2009}. 
The flat spectrum at around \( f = 1 - 10^3\) Hz is a direct consequence of the time variability of the gamma-ray emission. A GRB with \(E_{\gamma} = 10^{52}-10^{53}\) erg and \(\Delta \theta = 0.1 \) 
at \(d = 10 \) kpc (Galactic Center) is likely to provide an amplitude at a level close to the sensitivity
of Advanced LIGO detector at around 100 Hz (\cite{Flanagan1998}).
Thus the GW from GRB jets may be an interesting target for Einstein Telescope (ET)  (\cite{ETweb}) as well as for Advanced LIGO.
  
Unfortunately, the local GRB event rate has been estimated to be relatively small
as \(\sim 1 \) Gpc\(^{-3}\) yr\(^{-1}\) (\cite{Matsubayashi2005}; \cite{Wanderman2010}). 
This translates into an event rate \(\sim 10^{-9}\) yr\(^{-1}\) ( within \(d=1\) Mpc ) or \(\sim 10^{-6}\) 
yr\(^{-1}\) ( within \(d=10\) Mpc ) as a rate of GRBs with detectable GW memory. 
The possibility of detecting the GW memory 
from on-axis viewing angles, that is, almost simultaneously with gamma-rays from GRBs, makes it easier
to extract the memory component embedded from the detector's signals.  
Thus the GW memory from GRB jets will be one of the interesting targets for the next generation 
detectors such as KAGRA (\cite{Somiya2012}), DECIGO/BBO, and ET.  

\bigskip

The authors would like to thank the anonymous referee for valuable comments and suggestions, which greatly improved the manuscript.
K.I. is grateful to Drs. Takashi Nakamura, Kunihito Ioka, and Kenji Toma for stimulating discussion. 

\bigskip

%In figure \ref{fig:sample}, ...

%In the theory (\cite{key-1})..........

% In the theory (\cite{key-2})..........

% \subsubsection{Subsubsection}

% The resent result from ...........

% \newpage

% \section{Section 4}

% The final ..........

%%%%%%%%%%%%%%%%%%%%%%%%%%%%%%%%%%%%%%%

\bigskip

%Acknowledgement should be placed at end of main text.
%(NOT after the Appendix.)

%\appendix
%\section{Method of .....}

%%%
% See the manual for the detail.
%%%

\end{document}